\documentclass[journal]{IEEEtran}
\ifCLASSINFOpdf
\else
\fi
\usepackage[cmex10]{amsmath}
\usepackage{array}
\usepackage{mdwmath}
\usepackage[font=footnotesize]{subfig}
\usepackage{url}
\usepackage{amsmath,dsfont}
\usepackage{amssymb}
\usepackage{tikz}
\usepackage{caption}
\usepackage{algorithm}
\usepackage{algpseudocode}
\usepackage{pgfplots}
\usepackage{multirow}
\usepackage{enumitem}
\usepackage{comment}
\usepackage{graphicx}

\newtheorem{claim}{Claim}

\newtheorem{lemma}{Lemma}

\newtheorem{definition}{Definition}
\newtheorem{remark}{Remark}
\newtheorem{example}{Example}

\algnewcommand{\Inputs}[1]{%
	\State \textbf{Inputs:}
	\Statex \hspace*{\algorithmicindent}\parbox[t]{.8\linewidth}{\raggedright #1}
}
\algnewcommand{\Initialize}[1]{%
	\State \textbf{Initialize:}
	\Statex \hspace*{\algorithmicindent}\parbox[t]{.8\linewidth}{\raggedright #1}
}
\algnewcommand{\Output}[1]{%
	\State \textbf{Outputs:}
	\Statex \hspace*{\algorithmicindent}\parbox[t]{.8\linewidth}{\raggedright #1}
}

\newcommand{\bproof}{ \begin{IEEEproof} }
	\newcommand{\eproof}{ \end{IEEEproof} }
\newcommand{\beqno}{ \begin{equation*} }
\newcommand{\eeqno}{ \end{equation*} }
\newcommand{\beqa}{\begin{eqnarray*} }
	\newcommand{\eeqa}{\end{eqnarray*} }
\newcommand{\beq}{ \begin{equation} }
\newcommand{\eeq}{ \end{equation} }

\newcommand{\calM}{\mathcal{M}}
\newcommand{\calC}{\mathcal{C}}

\newcommand{\calN}{\mathcal{N}}

\newcommand{\calI}{\mathcal{I}}
\newcommand{\calU}{\mathcal{U}}

\newcommand{\calV}{\mathcal{V}}
\newcommand{\calX}{\mathcal{X}}

\newcommand{\calF}{\mathcal{F}}

\newcommand{\B}{\mathbf{B}}
\newcommand{\hatU}{\hat{U}}

\newcolumntype{L}[1]{>{\raggedright\arraybackslash}p{#1}}
\newcolumntype{C}[1]{>{\centering\arraybackslash}p{#1}}
\newcolumntype{R}[1]{>{\raggedleft\arraybackslash}p{#1}}

\usepackage{cite}

\begin{document}
	%
	\title{Asynchronous Coded Caching with Uncoded Prefetching}

\author{Hooshang~Ghasemi
        and~Aditya~Ramamoorthy
\thanks{ This work was supported in part by the
National Science Foundation (NSF) under grants CCF-1718470 and CCF-1910840. The material in this work has appeared in part at the 2017 IEEE International Symposium on Information Theory and the 2017 Asilomar Conference on Signals, Systems and Computers. Hooshang Ghasemi was with Iowa State University, he is now with Qualcomm Inc. Aditya Ramamoorthy is with Iowa State University, Ames, IA, 50011 USA. E-mail: \{hghasemi@qti.qualcomm.com, adityar@iastate.edu\}.
}
}


	\maketitle
	
	\begin{abstract}
		Coded caching is a technique that promises huge reductions in network traffic in content-delivery networks.
		However, the original formulation and several subsequent contributions in the area, assume that the file requests from the users are synchronized, i.e., they arrive at the server at the same time.
		In this work, we formulate and study the coded caching problem when the file requests from the users arrive at different times. We assume that each user also has a prescribed deadline by which they want their request to be completed. In the offline case, we assume that the server knows the arrival times before starting transmission and in the online case, the user requests are revealed to the server over time. We present a linear programming formulation for the offline case that minimizes the overall transmission rate from the server subject to the constraint that each user meets his/her deadline. While the online case is much harder,
		we introduce a novel heuristic for it and show that under certain conditions, with high probability the request of each user can be satisfied with her/his deadline. Our simulation results indicate that in the presence of mild asynchronism, much of the benefit of coded caching can still be leveraged.
	\end{abstract}
	
	\begin{IEEEkeywords}
		coded caching, asynchronous, deadlines, linear programming
	\end{IEEEkeywords}
	\IEEEpeerreviewmaketitle
	
	\section{Introduction}

Caching is a core component of solving the problem of large scale content delivery over the Internet. Conventional caching typically relies on placing popular content closer to end-users. Statistically, popular content is requested more frequently and the cache can be used to serve the user requests in this case. Contacting the central server that has all the content is not needed. This serves to reduce the induced network traffic.

%
	
In their pioneering work \cite{maddahN14}, Maddah-Ali and Niesen considered the usage of coding in the caching problem. In this ``coded caching" setting, there is a server containing a library of $N$ files. There are $K$ users each with a cache that can store up to $M$ files. The users are connected to the server via an error-free shared broadcast link (see Fig. \ref{Fig:block_coded_caching}).
	The system operates in two distinct phases. In the \textit{placement phase} the content of the caches is populated by the server. This phase does not depend on the future requests of the users which are assumed to be arbitrary. In the \textit{delivery phase} each user makes a request and the server transmits potentially coded signals to satisfy the requests of the users. The work of \cite{maddahN14} demonstrated that significant reductions in the network traffic were possible as compared to conventional caching. Crucially, these gains continue to hold even if the popularity of the files is not taken into account.

	
	While this is a significant result, the original formulation of the coded caching problem assumes that the user requests are synchronized, i.e., all file requests from the users arrive at the server at the same time. Henceforth, we refer to this as the synchronous setting. From a practical perspective, it is important to consider the asynchronous setting where user requests arrive at different times.  
	In this case, a simple strategy would be to wait for the last request to arrive and then apply the scheme of \cite{maddahN14}. Such a strategy will be quite good in terms of the overall rate of transmission from the server. However, this may be quite bad for an end user's experience, e.g., the delay experienced by the users will essentially be dominated by the arrival time of the last request.
	
	In this work, we formulate and study the coded caching problem when the user requests arrive at different times. Each user has a specific deadline by which his/her demand needs to be satisfied. The goal is to schedule the transmission of packets so that each user is able to recover the requested file from the transmitted packets and his/her cache content within the prescribed deadline. We present algorithms for both the offline and online versions of this problem.

	
This paper is organized as follows. In Section \ref{sec:backgd_rel_work_contrib} we discuss the background and related work and overview our main contributions.
The problem formulation appears in Section \ref{sec:problem_formulation}. Sections \ref{sec:offline} and \ref{sec:online_sec} discuss our work on the offline and the online versions of the problem, respectively.
We conclude the paper with a discussion of opportunities for future work in Section \ref{sec:conclusion}.


\vspace{-0.1in}
\section{Background, Related Work and Summary of Contributions}
\label{sec:backgd_rel_work_contrib}
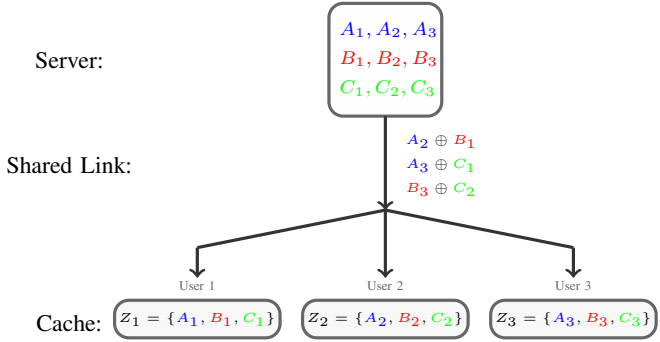
\begin{figure}[!t]
	\begin{center}
		\begin{tikzpicture}
		\shade[top color=white!30]
		[draw=black!60,fill=black!20,rounded corners=1.2ex,very thick] (-0.75,0.25) rectangle (0.75,1.75);
		\node[align=left] at (-4.2,1) {\small{Server:}};
		\node[align=left] at (-4.2, -0.4) {\small{Shared Link:}};
		\node[align=left] at (-4.2, -2.5) {\small{Cache:}};
		\node[color=black!80](server) at (0.2, 1) {\scriptsize{$\begin{aligned}
				&\textcolor{blue}{A_1,A_2,A_3}&\\
				&\textcolor{red}{B_1,B_2,B_3} &\\
				&\textcolor{green}{C_1,C_2,C_3}&
				\end{aligned}$}};
		\draw[->,very thick, black!80] (0,0.25) -- (0,-1) node[label={[xshift=0.9cm, yshift=-0.1cm]{\tiny{$\begin{aligned}
					&\textcolor{blue}{A_2}\oplus\textcolor{red}{B_1} &\\
					&\textcolor{blue}{A_3}\oplus\textcolor{green}{C_1} &\\
					&\textcolor{red}{B_3}\oplus\textcolor{green}{C_2}&
					\end{aligned}$}}}] {};
		\draw[very thick, black!80] (0,-1) -- (-2.5,-1.5);
		\draw[->,very thick, black!80] (-2.5,-1.5) -- (-2.5,-1.9);
		\draw[->, very thick, black!80] (0,-1) -- (0,-1.9);
		\draw[very thick, black!80] (0,-1) -- (2.5,-1.5);
		\draw[->,very thick, black!80] (2.5,-1.5) -- (2.5,-1.9);
		
		\node[color=black!60](usr1) at (-2.5,-2) {\tiny{User 1}};
		\node[color=black!60](usr3) at (0,-2) {\tiny{User 2}};
		\node[color=black!60](usr4) at (2.5, -2) {\tiny{User 3}};
		
		\shade[top color=white!30]
		[draw=black!60,fill=black!20,rounded corners=1.2ex,very thick] (-3.6,-2.7) rectangle (-1.4,-2.2);
		\node[color=black](server) at (-2.5, -2.45) {\tiny{$Z_1 = \{\textcolor{blue}{A_1},\textcolor{red}{B_1},\textcolor{green}{C_1}  \}$
		}};
		\shade[top color=white!30]
		[draw=black!60,fill=black!20,rounded corners=1.2ex,very thick] (-1.1,-2.7) rectangle (1.1,-2.2);
		\node[color=black](server) at (0, -2.45) {\tiny{$Z_2 = \{\textcolor{blue}{A_2},\textcolor{red}{B_2},\textcolor{green}{C_2}  \}$
		}};
		\shade[top color=white!30]
		[draw=black!60,fill=black!20,rounded corners=1.2ex,very thick] (1.4,-2.7) rectangle (3.6,-2.2);
		\node[color=black](server) at (2.5, -2.45) {\tiny{$Z_3 = \{\textcolor{blue}{A_3},\textcolor{red}{B_3},\textcolor{green}{C_3}  \}$
		}};
		;
		\end{tikzpicture}
	\end{center}
	\caption{\label{Fig:block_coded_caching} {\small Block diagram of the coded caching system with $N=K=3$ where cache of each user contains $1/3$-rd fraction of each file as shown. Users 1, 2, and 3 request file $A$, $B$, and $C$ respectively and the server transmits three subfiles over the shared link to satisfy their demands.}}
\vspace{-0.1in}
\end{figure}

%
A coded caching system contains a server with $N$ files, denoted $W_n$, $n =1,\ldots,N$, each of size $F$ subfiles, where a subfile is a basic unit of storage. These subfiles are indexed as $W_{n,j}, j = 1, \dots, F$. The system also contains $K$ users each connected to the server through an error free, broadcast shared link.
Each of the users is equipped with a local cache. The $i$-th cache can store the equivalent of $M_iF$ subfiles. We denote the cache content of user $i$ by $Z_i$, where $Z_i$ is a function of $W_{1}, \ldots, W_N$. Our formulation supports users with different cache sizes. A block diagram of a coded caching system for $N=K=3$ is depicted in Fig. \ref{Fig:block_coded_caching}.

In general, a user might choose to store a coded combination of the subfiles, i.e., $Z_i$ can be a non-trivial function of $\{W_{n,j}\}_{n=1,\dots,N, j= 1, \dots, F}$.
However, in this work, we assume that an uncoded placement scheme is being used by the coded caching system, i.e., user $i$ caches at most a $M_iF$-sized subset of all the subfiles at the server, i.e., $Z_i$ is a subset of $\{W_{n,j}\}_{n=1,\dots,N, j= 1, \dots, F}$. The uncoded placement scheme was shown in \cite{maddahN14} to have excellent performance. It has also been considered in several follow-up works as well. It is well recognized that the delivery phase in the uncoded placement case, corresponds to an index coding problem \cite{arbabjolfaei2018fundamentals}. While the optimal solution for an arbitrary index coding problem is known to be hard, techniques such as clique cover on the side information graph are well-recognized to have good performance \cite{arbabjolfaei2018fundamentals}. 
In this case, each transmitted equation from the server is such that a certain number of users ``benefit" from it simultaneously. Under this assumption, we formulate and study the asynchronous coded caching problem when the file requests arrive at the server at different times. Each user specifies a deadline by which he/she expects the request to be satisfied\footnote{It is not too hard to see that in the absence of deadlines the server can simply wait for enough user requests to arrive before starting transmission. Thus, the deadline-free case essentially reduces to the synchronous setting. Section IV.B of \cite{ghasemi2017asynch} has more details.}. We assume that
\begin{itemize}[wide, labelwidth=!, labelindent=0pt]
	\item the delivery phase proceeds via a clique cover and
	\item transmitting a single packet over the shared link takes a certain number of time slots.
\end{itemize}
We study the rate gains of coded caching under this setup, i.e., among the class of strategies that allow the users to meet their deadlines, we attempt to determine those where the server transmits the fewest number of packets. Both the offline and online versions of the problem are studied. In the offline scenario, we assume that information about all request arrival times and deadlines are known to the server before transmission, whereas in the online scenario, the arrival times and deadlines are revealed to the server as time progresses.


\subsection{Main contributions}

\begin{itemize}[wide, labelwidth=!, labelindent=0pt]
	\item {\it Linear programming (LP) formulation in the offline case.} We propose an LP in the offline scenario that determines a schedule for the equations that need to be transmitted from the server. A feasible point of the LP can be interpreted as a coding solution that can be used by the server, such that each user meets its deadlines. 

The computational complexity of solving this LP can be quite high for a large number of users. Accordingly, we develop a dual decomposition technique where the dual problem decouples into a set of independent minimum cost network flow problems that can be solved efficiently \cite{rk}. 

\item {\it A novel online algorithm.} For the online problem, we demonstrate that, in general, coding within subfiles of the same file is essential.
Interestingly, this is not needed in the synchronous case where the transmitted signals by the server are coded combination of subfiles belonging to different files.
Furthermore, we propose a novel online algorithm that is inspired by recursively solving the offline LP and interpreting the corresponding output appropriately. Under certain conditions, we also show that the algorithm will result in a solution that satisfies the deadline constraints with high probability. 

	
\end{itemize}
For both scenarios, we present exhaustive simulation results that corroborate our findings and demonstrate the superiority of our algorithm concerning prior work.
Overall, our work indicates that under mild asynchronism, much of the benefits of coded caching can still be leveraged.

\subsection{Related Work}
The area of coded caching has seen a flurry of research activity along several dimensions in recent years. From a theoretical perspective, significant work has attempted to understand the fundamental rate limits of a coded caching system  \cite{ghasemiR17_jnl,yu2017exact,QYuMA19}. Extensions of the basic model to general networks have been examined in \cite{TangR16,wanJPT18,naderializadeh2017fundamental}. Issues related to subpacketization (i.e., the number of subfiles $F$) have been considered in \cite{yanCTC17,tang2018coded,lampirisE18}. A high subpacketization level can cause several issues in practical implementations. Coded caching ideas have also been used within the domain of distributed computing \cite{li2016fundamental, kiamari2017heterogeneous, kostasR20}.


There are relatively few prior works that have considered asynchronism within the context of coded caching. To our best knowledge, it was first studied in \cite{niesen2015coded}. They considered the decentralized coded caching model \cite{maddah2015decentralized} and a situation where each subfile has a specific deadline. Only the online case was considered and heuristics for transmission from the server were proposed. The heuristics are found to have good performance. However, the transmission time for a packet was not considered in their formulation. Reference \cite{luCP18} also considers the asynchronous setting; again, they do not consider the transmission time of a transmitted packet. In that sense, their setting is closer to the work of \cite{niesen2015coded} and can be viewed as a set of rules that the server should follow in the online case. \cite{luCP18} (Section III.C) also considers an offline setting for the centralized placement scheme of \cite{maddahN14}.
Our LP formulation can be viewed as a bound on the possible performance of any online scheme. Our proposed online algorithm has significantly better performance than the ones presented in \cite{niesen2015coded}.


Reference \cite{maddah2015decentralized} (Section V.C) also discusses the issue of asynchronism within the context of decentralized coded caching, without considering deadlines or packet transmission times. They advocate a further subpacketization of each subfile (referred to as a segment in  \cite{maddah2015decentralized}). It is important to note that any system will need to commit to a certain subpacketization scheme before deployment. Given this subpacketization and with user specified deadlines, the formalism of our work and our algorithms can be used to arrive at schemes that address asynchronous requests.

The work of \cite{jiangHBZ19} proposes an algorithm for the online scenario under the assumption of decentralized coded caching for reducing the worst-case load of fronthaul links in fog radio access networks (F-RANs); this is a different model than ours. Their work does not take transmission time into account and considers the scenario where each user has the same deadline.

The asynchronous setting has also been considered in \cite{yang2018audience} for video delivery by taking into account an appropriately defined audience retention rate. Their work considers a probabilistic arrival model and presents a decentralized coded caching scheme for it.



\vspace{-0.1in}
\section{Problem Formulation and Preliminaries}
\label{sec:problem_formulation}
	
We assume that time $\tau \geq 0$ is slotted. Let $[n]$ denote the set $\{1, \ldots, n\}$ and the symbol $\oplus$ represent the XOR operation. We assume that the server contains $N\geq K$ files\footnote{We assume that $N\geq K$ as it corresponds to the worst case rate (under most reasonable placement schemes) where each of the $K$ users can request a different file. Furthermore, it is also the more practical scenario.} denoted by $W_{n}, n = 1, \dots, N$. The subfiles are denoted by $W_{n,f}$ so that $W_n = \{W_{n,f}: f \in [F]\}$ and the cache of user $i$ by $Z_i \subseteq \{ W_{n,f}: \ n \in [N], \ f \in [F] \}$. $Z_i$ contains at most $M_iF$ subfiles. In the delivery phase, user $i$ requests file $W_{d_i}$, where $d_i \in [N]$, from the server.
We let $\Omega^{(i)}$ denote the indices of the subfiles that are not present in the $i$-th user's cache, i.e.,
	\begin{align*}
	\Omega^{(i)} \triangleq \{ f:\ f \in [F], \ W_{d_i,f} \notin Z_i \}.
	\end{align*}
The equations in the delivery phase are assumed to be of the {\it all-but-one} type.
\begin{definition} {\it All-but-one equation.} Consider an equation $E$ such that
\begin{align*}
E \triangleq \oplus_{l=1}^\ell W_{d_{i_l}, f_{l}}.
\end{align*}
We say that $E$ is of the all-but-one type if for each $l \in [\ell]$, we have $W_{d_{i_l}, f_{l}} \notin Z_{i_l}$ and $W_{d_{i_l}, f_{l}} \in Z_{i_{k}}$ for all $k \in [\ell] \setminus \{l\}$.
\end{definition}
It is evident that an all-but-one equation transmitted from the server allows each of the users participating in the equation to recover a missing subfile that they need. The asynchronous coded caching problem can be formulated as follows.
	
\begin{figure}[t]
	\centering
	\begin{tikzpicture}
	\draw[thick, <->] (0,0.5) --(0.5*1.35,0.5) node[above] {\textcolor{black}{$\Pi_1$}} -- (1.35,0.5);
	\draw[thick, <->] (1.36,0.5) --(2*1.35,0.5) node[above] {\textcolor{black}{$\Pi_2$}} -- (3*1.35,0.5);
	\draw[thick, <->] (3*1.35-0.01,0.5) --(4*1.35,0.5) node[above] {\textcolor{black}{$\Pi_3$}} -- (5*1.35,0.5);
	
	\draw[thick, ->] (0,0) -- (0.5*1.35,0) node[above] {\tiny{$\textcolor{red}{W_{1,3}}$}} -- (1.5*1.35,0) node[above] {\tiny{$\textcolor{red}{W_{1,1}}\oplus \textcolor{blue}{W_{2,1}}$}}-- (2.5*1.35,0) node[above] {\tiny{$\textcolor{red}{W_{1,2}}$}}
	-- (3.5*1.35, 0) node[above]
	{\tiny{$\textcolor{blue}{W_{2,2}}\oplus \textcolor{green}{W_{3,1}}$}} -- (4.5*1.35,0) node[above] {\tiny{$\textcolor{green}{W_{3,3}}$}} --(6*1.35,0) node[right] {$\tau$};
	\draw [thick,
	postaction={
		draw,
		decoration=ticks,
		segment length=1.35cm,
		decorate,
	}
	] (0,0) -- (8,0);
	\foreach \tick in {0,...,5}
	\node at (1.35*\tick,0) [below=1pt] {\tick};
	
	\draw[thick, <->, color=red!80] (0,-0.5) node[below] {\textcolor{black}{$T_1$}} -- (5*1.35,-0.5);
	\draw[thick, <->, color=blue!80] (1*1.35,-0.75) node[below] {\textcolor{black}{$T_2$}}-- (5*1.35,-0.75);
	\draw[thick, <->, color=green!80] (3*1.35,-1)node[below] {\textcolor{black}{$T_3$}} -- (5*1.35,-1);
	
	\end{tikzpicture}

	\caption{\label{Fig:N_3_K_3_M_1_offline} {\small Offline solution corresponding to the Example \ref{eg:consolidated_eg_sec3} (system with $N=K=3$). The double-headed arrows show the active time slots for each user. The transmitted equations are shown above the timeline. }}
	\vspace{-0.2in}
\end{figure}
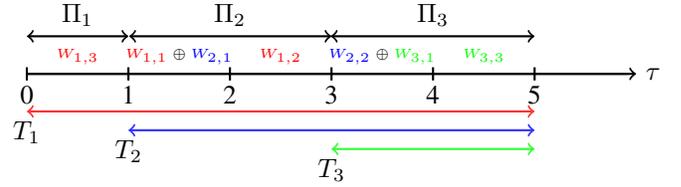

	\noindent {\it Inputs.}
	\begin{itemize}[wide, labelwidth=!, labelindent=0pt]
		\item {\it User requests.} User $i$ requests file $W_{d_i}$, with $d_i \in [N]$ at time $T_i$. 
		\item {\it Deadlines.} The $i$-th user needs to be satisfied by time $T_i + \Delta_i$, where $\Delta_i$ is a positive integer.
		\item {\it Transmission delay.} Each subfile needs $r$ time-slots to be transmitted over the shared link, i.e., each subfile can be treated as equivalent to $r$ packets, where each packet can be transmitted in one time slot.
	\end{itemize}
	As the problem is symmetric with respect to users, w.l.o.g. we assume that $T_1 \leq T_2 \leq \ldots \leq T_K$. Let $T_{\max} = \max_i (T_i + \Delta_i)$. Note that upon sorting the set of arrival times and deadlines, i.e., $\cup_{i=1}^{K} \{T_i, T_i + \Delta_i\}$, we can divide the interval $[T_1, T_{\max})$ into {\it at most} $2K-1$ non-overlapping intervals. Let the integer $\beta$, where $1 \leq \beta \leq 2K-1$ denote the number of intervals.
	Let $\Pi_1, \ldots, \Pi_{\beta}$ represent the intervals where $\Pi_i$ appears before $\Pi_j$ if $i < j$; $|\Pi_\ell|$ denotes the length of interval $\Pi_\ell$. The intervals are left-closed and right-open.  An easy to see but very useful property of the intervals that we have defined is that for a given $i$, either $[T_i, T_i + \Delta_i) \cap \Pi_\ell = \Pi_\ell$ or $[T_i, T_i + \Delta_i) \cap \Pi_\ell = \emptyset$. Fig. \ref{Fig:N_3_K_3_M_1_offline} shows an example when $K=3$. We define $U_\ell \triangleq \{i \in [K]:~ [T_i , T_i + \Delta_i) \cap \Pi_\ell = \Pi_\ell \}$, and
	$D_\ell \triangleq \{d_i\in [N]:~i\in U_\ell \}$.
Thus, $U_\ell$ is the set of active users in time interval $\Pi_\ell$ and $D_\ell$ is the corresponding set of active file requests.
	
\noindent {\it Outputs.}
	\begin{itemize}[wide, labelwidth=!, labelindent=0pt]
		\item {\it Transmissions at each time slot.} If the problem is feasible, the schedule specifies which equations (of the all-but-one type) need to be transmitted at each time. The schedule is such that each user can recover all its missing subfiles within its deadline. The equations transmitted at time $\tau \in \Pi_\ell$ only depend on $D_\ell$. 
	\end{itemize}
\noindent We consider two versions of the above problem.
\begin{itemize}[wide, labelwidth=!, labelindent=0pt]
\item {\it Offline version.} In the offline version, we assume that the server is aware of $\{T_i, \Delta_i, d_i\}_{i=1}^K$ at $\tau = 0$. However, at time $\tau \in \Pi_\ell$ the transmitted equation(s) will only depend on $D_\ell$, i.e., the server cannot start sending missing subfiles for a given user until its request arrives.
\item {\it Online version.} In the online version, information about the file requests are revealed to the server as time progresses. At each time $\tau$, the server only has information about $\{T_i, \Delta_i, d_i\}$ if $T_i \leq \tau$, i.e., the requests that have arrived by time $\tau$.
\end{itemize}

We begin by defining some relevant sets; for convenience, a tabulated list of most of the items needed in the subsequent sections can be found in Table \ref{Tab:Variable_Description}. Consider a subset of users $U \subseteq [K]$. For each user $i\in U$ we let $\calF_{\{i,U\}}$ denote the indices of all missing subfiles of the $i$-th user that have been stored in the cache of the other users in $U$, i.e.,
\begin{align*}
\calF_{\{i,U\}} \triangleq\Big\{f\in \Omega^{(i)}:\ W_{d_i,f} \in Z_j \text{ for all } j\in U\setminus \{i \} \Big\}.
\end{align*}
\begin{definition}
\label{defn:user_gp}
{\it User Group.} A subset $U \subseteq [K]$ is said to be a user group if $\calF_{\{i,U\}} \neq \emptyset$ for all users $i \in U$ so that there is at least one all-but-one type equation associated with $U$.
\end{definition}


For a user group $U$ there are $\prod_{i\in U} |\calF_{\{i,U\}}|$ different all-but-one equations. This is because for any choice of $f_i \in \calF_{\{i,{U}\}}$ for $i \in U$, we can construct the all-but-one equation $\oplus_{i\in U} W_{d_i,f_i}$. Thus, for each $i \in U$ there are $|\calF_{\{i,U\}}|$ choices for $f_i$.
%
%
Recall that $U_\ell$ is the set of active users in time interval $\Pi_\ell$ and $D_\ell$ represents their file requests. Let $\calU_\ell$ be a subset of the power set of $U_\ell$ (i.e. the set of all subsets of $U_\ell$) such that each element in $\calU_{\ell}$ is an user group ({\it cf.} Definition \ref{defn:user_gp}). For any $U \subseteq [K]$, let $\calI_U$ be the set of indices of all time intervals where the users in $U$ are simultaneously active, i.e.,
$$\calI_U \triangleq \Big\{\ell:~ [T_i ,T_i+\Delta_i) \cap \Pi_\ell = \Pi_\ell,~ \forall \ i \in U \Big\}.$$

For each missing subfile $W_{\{d_i,f\}}$ (where $f \in \Omega^{(i)}$) we let $\calU_{\{i, f\}}$ denote the set of user groups where it can be transmitted, i.e.,
$$\calU_{\{i, f\}} \triangleq\Big\{ U \in \cup_{\ell=1}^{\beta} \calU_\ell :\ i\in U, f \in \calF_{\{i,U\}} \Big\}.$$
We note here that for a fixed $i$, there are potentially multiple indices $f_1, f_2, \dots, f_l \in \Omega^{(i)}$ such that $U \in \calU_{\{i, f_j\}}$ for $j =1,\dots, l$. 

%
%

\begin{example}
\label{eg:consolidated_eg_sec3}
Consider a system (shown in Fig. \ref{Fig:N_3_K_3_M_1_offline}) with $N=3$ files, $W_1$, $W_2$, and $W_3$ where each file is divided into three subfiles, so that $F=3$. There are $K=3$ users with the following cache content, $Z_{1}	=	\{W_{2,1}, W_{2,2}, W_{3,3}\}$, $Z_{2}	=	 \{W_{1,1}, W_{2,3}, W_{3,1}\}$, and $Z_{3}	=	\{W_{2,2}, W_{3,2}, W_{2,1}\}$. Thus, $M_i=1$ for $i\in [K]$. The arrival times are $T_1=0$, $T_2=1$, $T_3=3$, and deadlines are $\Delta_1=5$, $\Delta_2=4$, and $\Delta_3=2$. The $i$-th user requests file $W_i$, for $i =1,\dots,3$. Therefore, $\Omega^{(1)} = \{1, 2, 3\}$, $\Omega^{(2)} = \{1,2\}$, and $\Omega^{(3)} = \{1,3\}$.

 In this system we have $\calF_{\{1,\{1,2\}\}} = \{1\}$ as $W_{1,1}\in Z_2$, and $\calF_{\{2,\{1,2\}\}} = \{1, 2\}$ as $W_{2,1},W_{2,2} \in Z_1$. Therefore, $\{1,2\}$ is an user group and the corresponding all-but-one equations are $W_{1,1} \oplus W_{2,1}$ and $W_{1,1} \oplus W_{2,2}$. However, $\calF_{\{1,\{1,3\}\}} = \emptyset$ thus $\{1,3\}$ is not an user group.

 As $U= \{1,2\}$ is an user group, we have $\calU_2 = \{\{1\}, \{2\}, \{1,2\}\}$.  The set of time intervals where user group $\{1,2\}$ is active is  $\calI_{\{1,2\}} = \{2,3\}$. Finally, note that user group $U = \{2,3\}$ is a member of $\calU_{\{2,1\}}$ since $2 \in U$ and $1 \in \calF_{\{2, U \}} = \{1,2\}$. Similarly, $U \in \calU_{\{2,2\}}$ as well since $2 \in U$ and $2 \in \calF_{\{2, U \}}$.

\end{example}

\vspace{-0.1in}
\section{Offline Asynchronous Coded Caching}
	\label{sec:offline}

In this section, we discuss the offline version of the problem where the server knows the arrival times/deadlines of all the requests at $\tau = 0$. 
The offline solution of the system in Example \ref{eg:consolidated_eg_sec3} is depicted in Fig. \ref{Fig:N_3_K_3_M_1_offline} where the transmitted equation in each time slot appears above the timeline. It can be verified that each user can recover the missing subfiles that they need. In what follows we argue that the offline setting can be cast as a linear programming problem.

\vspace{-0.1in}	
\subsection{Linear programming formulation}
\label{subsec:linprog}
	For each time interval $\Pi_\ell$ with $\ell = 1,\ldots, \beta$ and for each $U \in \calU_\ell$, we define variable $x_U(\ell) \in [0,|\Pi_\ell|]$ that represents the portion of time interval $\Pi_\ell$ that is allocated to an equation that benefits user group $U$. The actual equation will be determined shortly. For each missing subfile $W_{\{d_i,f\}}$ and each $U \in \calU_{\{ i,f\}}$, we define variable $y_{\{i,f\}}(U) \in [0,r]$ that represents the portion of the missing subfile $W_{\{d_i,f\}}$ transmitted within some or all of the equations associated with $x_U(\ell)$ for $\ell \in \calI_U$. As pointed out before, for a fixed $i$, $U$ can be used to transmit different missing subfiles needed by user $i$. However, a single equation can only help recover one missing subfile needed by $i$. Thus, $\sum_{\ell \in \calI_U} x_U(\ell)$ must be shared between the appropriate $y_{\{i,f\}}(U)$'s. Accordingly, we need the following constraint for user $i$ and a user group $U$ which contains $i$.
	\begin{align*}
	\sum_{f \in \calF_{\{i,U\}}} y_{\{i,f\}}(U) \leq \sum_{\ell \in \calI_U} x_U(\ell).
	\end{align*}
	In addition, at time interval $\Pi_\ell$ at most $|\Pi_\ell|$ packets can be transmitted, so that $\sum_{U \subseteq \calU_\ell} x_U(\ell) \leq |\Pi_\ell|$. To ensure that each missing subfile $W_{\{d_i,f\}}$ is transmitted in exactly $r$ time slots, we have $\sum_{U \in \calU_{\{ i, f\}}} y_{\{ i, f\}}(U) = r$.

\begin{table}[t]
	\centering
	\begin{tabular}{L{1.5cm}  L{6cm}}
		\hline
		Variable & Description \\ \hline
		$T_i$ & arrival time of user $i$ \\ \hline
		$T_i + \Delta_i$ & deadline of user $i$ \\ \hline
		$\beta$ & number of time intervals\\ \hline
		$\Pi_{\ell}$ & time interval $\ell$ \\ \hline
		$U_\ell$ & set of the active users in time interval $\ell$ \\ \hline
		$\Omega^{(i)}$ & set of the indices of missing subfiles of user $i$ \\ \hline
		$\calU_\ell$ & set of all subsets of $U_\ell$ that are user groups \\ \hline
		$\calI_U$ & set of the indices $\ell \in [\beta]$ so that $U \in \calU_\ell$ \\ \hline
		$\calU_{\{ i, f\}}$ & set of all user groups that $W_{d_i,f}$ can be transmitted within \\ \hline
		$x_U(\ell)$ & portion of $\Pi_\ell$ allocated to user group $U$ \\ \hline
		$y_{\{i,f\}}(U)$ & portion of $W_{d_i,f}$ transmitted within user group $U$ \\ \hline
		$\calF_{\{i,U\}}$ & set of the indices of $f \in \Omega^{(i)}$ that can be transmitted within $U$ \\ \hline
	\end{tabular}
\caption{\label{Tab:Variable_Description}{\small List of variables used in the description}\vspace{-2mm}}
\end{table}

	
	The following LP minimizes the overall rate of transmission from the server while respecting all the deadline constraints of the users under the assumption that the server only transmits all-but-one equations. However, we point out that in general this may not be the information-theoretically optimal strategy for the server.
	\begin{align}
	\label{eq:simplfied_linear_program}
	\min_{\{x_U(\ell), y_{\{i,f \}}(U)\}} & \sum_{\ell=1}^{\beta} \sum_{U \in \calU_\ell} x_{U}(\ell) & \\
	\text{s.t. }~\sum_{U \in \calU_\ell} x_U(\ell) &\leq |\Pi_\ell|, \qquad \text{for }  \ell=1,\ldots,\beta,&  \nonumber\\
	\sum_{f \in \calF_{\{i,U\}}}y_{\{i,f\}}(U) &\leq \sum_{\ell \in \calI_U} x_{U}(\ell),~\text{for }  i \in U, \  U \in \cup_{\ell=1}^\beta \calU_\ell ,&\nonumber\\
	\sum_{U \in \calU_{\{i,f \}}} y_{\{i,f\}}(U)&= r,  \qquad \text{for } f \in \Omega^{(i)},\ i \in [K],  & \nonumber\\
	 x_U(\ell) ,~ y_{\{i,f \}}(U)&\geq 0, \qquad \text{for } \forall i \in [K],\ \ell \in [\beta], \  U \in \cup_{\ell=1}^\beta \calU_\ell.& \nonumber
	\end{align}
Note that \cite{niesen2015coded} considers the case when each missing subfile has a prescribed deadline. Our LP above can be modified in a straightforward manner to incorporate this aspect.

\vspace{-0.1in}
\subsection{Interpretation of feasible point of (\ref{eq:simplfied_linear_program}) as a coding solution}
	\label{subsec:LP_interper}

We start by assigning time intervals to user groups. The time interval $\Pi_\ell$, $\ell \in [\beta]$, will be arbitrarily assigned to user groups $U \in \calU_\ell$ so that the time assigned to one user group does not overlap with another. The constraint $\sum_{U\in \calU_\ell}x_U(\ell) \leq |\Pi_{\ell}|$ implies that such an assignment exists.
For each user group $U$ and each $i \in U$, suppose that $f_1,\ldots,f_l \in \calF_{\{i,U\}}$ are such that $y_{\{i,f_{j}\}}(U) \neq 0$ for $j = 1,\ldots, l$. We assign $y_{\{i,f_j\}}(U)$ part of the total time allocated to user group $U$, i.e., $\sum_{\ell \in \calI_U} x_{U}(\ell)$, to the missing subfile $W_{d_i,f_j}$ for $j = 1,\ldots,l$. The constraint $\sum_{f \in \calF_{\{i,U\}}}y_{\{i,f\}}(U) \leq \sum_{\ell \in \calI_U} x_{U}(\ell)$ ensures that such an assignment always exists, i.e., it is possible to assign $y_{\{i,f\}}(U)$'s (for fixed $i$) to the available (strictly) positive $x_{U}(\ell)$'s, such that there is no overlap between them. This assignment is not unique in general. However, this is not a problem as any assignment can be used to determine the equations. This process is repeated for all users $i \in U$.
	
The equation transmitted on a particular interval is simply the XOR of the subfile indices that map to that interval. 
This equation is valid since the missing subfile $W_{d_i,f}$ with $f \in \calF_{\{i,U\}}$ is in the cache of all the users in $U\setminus \{i\}$.

Finally, according to the constraint $\sum_{U \in \calU_{\{i,f \}}} y_{\{i,f\}}(U)= r$, each missing subfile $W_{d_i,f}$ is transmitted in its entirety in some equations. The following example serves to illustrate the arguments above.
	
	\usetikzlibrary{decorations.pathreplacing,angles,quotes}
	\begin{figure}[t]
		\centering
		\scalebox{0.73}{
		\begin{tikzpicture}
		
\draw[thick, <->] (0,-0.5) --(1,-0.5) node[below] {\textcolor{black}{$\Pi_1$}} -- (1.99,-0.5);
\draw[thick, <->] (2,-0.5) --(4,-0.5) node[below] {\textcolor{black}{$\Pi_2$}} -- (5.99,-0.5);
\draw[thick, <->] (6,-0.5) --(8,-0.5) node[below] {\textcolor{black}{$\Pi_3$}} -- (9.99,-0.5);%
\draw [thick,
postaction={
	draw,
	decoration=ticks,
	segment length=2cm,
	decorate,
}
] (0,0) -- (10.01,0);
\foreach \tick in {0,...,5}
\node at (2*\tick,0) [below=1pt] {\tick};

\draw[line width=5, color=red!20] (2,0.1) -- (3,0.1);
\draw[line width=5, color=red!20] (8,0.1) -- (9,0.1);
\draw[line width=5, color=blue!20] (6,0.1) -- (7.99,0.1);
\draw[line width = 3, color=red!80] (2,0.3) -- (3,0.3);
\draw[line width=3, dotted, color=red!80] (8,0.3) -- (9,0.3);
\draw[line width=3, color=blue!80] (6,0.3) -- (6.99,0.3);
\draw[line width=3, dotted, color=blue!80] (7,0.3) -- (7.99,0.3);
\draw[line width=3, color=gray!80] (6,0.5) -- (8,0.5);
\draw[line width=3, color=green!80] (2,0.5) -- (3,0.5);
\draw[line width=3,color=green!80] (8,0.5) -- (9,0.5);

\draw[thick, <->] (2,0.7) --(2.5,0.7) node[above] {\textcolor{black}{$E_1$}} -- (3,0.7);
\draw[thick, <->] (6,0.7) --(6.5,0.7) node[above] {\textcolor{black}{$E_2$}} -- (7,0.7);
\draw[thick, <->] (7,0.7) --(7.5,0.7) node[above] {\textcolor{black}{$E_3$}} -- (8,0.7);
\draw[thick, <->] (8,0.7) --(8.5,0.7) node[above] {\textcolor{black}{$E_4$}} -- (9,0.7);

\draw[line width=5, color=red!20] (0,-1.5) -- (0.5,-1.5) node[right] {\textcolor{black}{ time assigned to user group $\{1,2\}$}};
\draw[line width=5, color=blue!20] (6,-1.5) -- (6.5,-1.5) node[right] {\textcolor{black}{ time assigned to user group $\{2,3\}$}};
\draw[line width=5, color=blue!80] (0,-2) -- (0.5,-2) node[right] {\textcolor{black}{ time assigned to $y_{\{2,1\}}(\{2,3\})$}};
\draw[line width=5, dotted, color=blue!80] (6,-2) -- (6.5,-2) node[right] {\textcolor{black}{ time assigned to $y_{\{2,2\}}(\{2,3\})$}};
\draw[line width=5, color=red!80] (0,-2.5) -- (0.5,-2.5) node[right] {\textcolor{black}{ time assigned to $y_{\{2,1\}}(\{1,2\})$}};
\draw[line width=5, dotted, color=red!80] (6,-2.5) -- (6.5,-2.5) node[right] {\textcolor{black}{ time assigned to $y_{\{2,2\}}(\{1,2\})$}};
\draw[line width=5, color=green!80] (0,-3) -- (0.5,-3) node[right] {\textcolor{black}{ time assigned to $y_{\{1,1\}}(\{1,2\})$}};
\draw[line width=5, color=gray!80] (6,-3) -- (6.5,-3) node[right] {\textcolor{black}{ time assigned to $y_{\{3,1\}}(\{2,3\})$}};

		\end{tikzpicture}
		}
		
		\caption{\label{Fig:interpret_solution} {\small Interpretation of feasible point in (\ref{eq:simplfied_linear_program}) for Example \ref{eg:consolidated_eg_sec3}. For readability, only equations corresponding to user groups $\{1,2\}$ and $\{2,3\}$ are depicted.}\vspace{-3mm}}
	\end{figure}
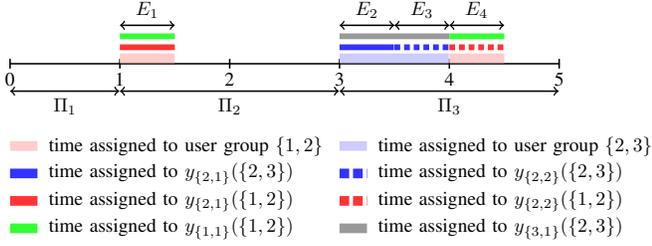
	
	\begin{example}
	\label{exm:assignment}Consider again the system in Example \ref{eg:consolidated_eg_sec3}. Part of a feasible solution to the LP in (\ref{eq:simplfied_linear_program}), corresponding to user groups $U = \{2,3\}$ and $U' = \{1,2\}$, is presented below.
	{\small
		\begin{align*}
		 x_{\{1,2\}}(2) = 0.5, ~~ & x_{\{1,2\}}(3)= 0.5,  &x_{\{2,3\}}(3) = 1, \\
		 y_{\{1,1\}}(\{1,2\}) = 1, ~~& y_{\{2,1\}}(\{1,2\}) = 0.5, & y_{\{2,2\}}(\{1,2\}) = 0.5, \\
		 y_{\{2,1\}}(\{2,3\}) = 0.5,~~ & y_{\{2,2\}}(\{2,3\}) = 0.5, & y_{\{3,1\}}(\{2,3\}) = 1.
		\end{align*}
    }
According to the solution, $x_{\{2,3\}}(3) =1$. Therefore, only one unit of $\Pi_3$ is assigned to $U$ (though $|\Pi_3| = 2$). This is denoted by the light blue color line in Fig. \ref{Fig:interpret_solution}. For user $3 \in U$, there is only one missing subfile in $\calF_{\{3,\{2,3\}\}}$, namely $W_{3,1}$. As $y_{\{3,1\}}(\{2,3\}) = 1$ it is assigned to  $x_{\{2,3\}}(3)$ in its entirety. This is depicted by the gray line in Fig. \ref{Fig:interpret_solution}. For user 2 in $U$ we have $\calF_{\{2,\{2,3\}\}} = \{1,2\}$. The solution specifies $y_{\{2,1\}}(\{2,3\}) = y_{\{2,2\}}(\{2,3\}) = 0.5$. Thus, we assign the first half of $x_{\{2,3\}}(3)$ to missing subfile $W_{2,1}$ and the second half to $W_{2,2}$ (see the dark blue and dotted dark blue lines in Fig. \ref{Fig:interpret_solution}).
%
Accordingly, the server transmits equations such that the first half of the time interval assigned to user group $U$ corresponds to the $E_2 = W_{2,1} \oplus W_{3,1}$ whereas the second half corresponds to $E_3 = W_{2,2} \oplus W_{3,1}$. The interpretation of the user group $U'$ is similar (see Fig. \ref{Fig:interpret_solution}). 
	\end{example}
\begin{remark}
The output of the above LP will typically result in a fractional solution for the variables. A fractional solution can be interpreted by assuming that each packet that is transmitted over the shared edge can be subdivided as finely as needed. Thus, in each time slot, we could transmit multiple equations that may serve potentially different subsets of users. This assumption is reasonable if the underlying subfiles and hence the packets are quite large. In any case, the above LP provides a lower bound on the performance of a solution where integrality constraints are enforced.
\end{remark}
\begin{remark}
\label{remark:offline_ordering}
We note that for the offline solution, within a given time interval, the user groups can be assigned in any order according to the $x_U(\ell)'s$ as long as they don't overlap. Moreover, the assignment of $y_{i,f}(U)'s$ is also arbitrary as long as the constraints of the LP are respected. However, for the online case ({\it cf.} Section \ref{sec:online_sec}), the order does matter since we make the best effort decision on each individual slot as we do not know the future arrivals.


\end{remark}

\subsection{Modified LP with fixed user group assignments}
\label{sec:modified_LP_offline}
Note that the LP in (\ref{eq:simplfied_linear_program}) includes the variables $x_U(\ell)$'s that determine the user groups in the different time-intervals. Suppose instead that at a certain time $\tau$ we are given the total time allocated to user group $U$ thus far (denoted by $\tilde{z}_U$) and only need to determine the $y_{\{i,f \}}(U)$ values for each $i$. 
Let $\tilde{\calU}$ be the set of user groups used until time $\tau$. For user $i \in [K]$, this can be written as a related LP as shown below which returns the total number of missing packets that user $i$ has obtained until time $\tau$.
	\begin{align}
	\label{eq:fixed_u_linear_program}
	\max_{\{y_{i,f}(U)\}} & \sum_{U \in \tilde{\calU}} \sum_{f \in \mathcal{F}_{i, U}} y_{i,f}(U) & \\
	\text{s.t. }~\sum_{f \in \calF_{\{i,U\}}}y_{\{i,f\}}(U) &\leq \tilde{z}_U,~\text{for }  i \in U, \  U \in \tilde{\calU} ,&\nonumber\\	
	\sum_{U \in \calU_{\{i,f \}}} y_{\{i,f\}}(U) &\leq r,  \qquad \text{for } f \in \Omega^{(i)},  & \nonumber\\
	 y_{\{i,f \}}(U)&\geq 0, \qquad \text{for }  U \in \tilde{\calU}.& \nonumber
	\end{align}

This follows since given the $z_U$'s we only need to find an assignment of the $y_{\{i,f\}}(U)$'s to the corresponding $z_U$'s that respect the first constraint above. Moreover, since we are not considering the entire transmission time, each missing subfile may not be transmitted in its entirety.

For instance, in Example \ref{eg:consolidated_eg_sec3} suppose that at time $\tau=4$, we have $z_{\{1\}} =2, z_{\{1,2\}}=1, z_{\{2,3\}}=1$, then the LP above in (\ref{eq:fixed_u_linear_program}) for user 1 has the optimum point $y_{\{1,1 \}}(\{1\}) = y_{\{1,3 \}}(\{1\})=1$, $y_{\{1,2 \}}(\{1,2\}) = 1$. Likewise for user 2, the LP has the optimum point $y_{\{2,1 \}}(\{1,2\}) = 1$ and $y_{\{2,2 \}}(\{2,3\}) = 1$.

\begin{remark}
The complexity of our solution in (\ref{eq:simplfied_linear_program}) does not have any dependence on arrival times $T_i$'s and deadlines $\Delta_i$'s. Our formulation of the LP in terms of the intervals allows us to circumvent this potential dependence. 
\end{remark}
Nevertheless, the complexity of solving the LP does grow quite quickly (cubic) in the problem parameters (number of constraints + number of variables) \cite{vaidyaLP}. Next, we discuss a solution based on dual decomposition that is much faster.

\vspace{-0.1in}
\subsection{Dual Decomposition based LP solution}	
	\label{sec:Dual_Decomposition}
	
	As it stands, the LP in (\ref{eq:simplfied_linear_program}) cannot be interpreted as a network flow. Yet, intuitively one can view the missing subfiles from each user as flowing through the user groups and getting absorbed in sinks that correspond to their valid time intervals. However, the flows corresponding to different users can be shared as the all-but-one equations allow different users to benefit from the same equation. We note here that a similar sharing of flows also occurs in the problem of minimum cost multicast with network coding \cite{lun2006minimum}. The LP in (\ref{eq:simplfied_linear_program}) can, however, be modified slightly so that the corresponding dual function is such that it can be evaluated by solving a set of {\it decoupled minimum cost network flow optimizations}.

	
\subsubsection{Decoupling procedure}

For each user $i \in U$ the variable $x^{(i)}_U(\ell)$ represents the amount of flow corresponding to user $i$ outgoing from user group $U$ to time interval $\Pi_\ell$. Evidently, this amount can't be more than $x_U(\ell)$. Therefore, we have
	\begin{align*}
	x^{(i)}_U(\ell) \leq x_U(\ell),
	\end{align*}
	which holds for all $i \in U$ and all $U \in  \calU_\ell$, $\ell \in [\beta]$. We define $\calU_\ell^{(i)} \subseteq \calU_\ell$ to be the subset of possible user groups at time interval $\Pi_\ell$ that include user $i$, i.e., $i \in U$ for all $U \in \calU_\ell^{(i)}$. By the flow interpretation  of $x^{(i)}_U(\ell)$, we have
	\begin{align*}
	\sum_{f \in \calF_{\{i,U\}}}y_{\{i,f\}}(U) = \sum_{\ell \in \calI_U} x_{U}^{(i)}(\ell),
	\end{align*}
	 for all $U \in \cup_{\ell=1}^\beta \calU^{(i)}_\ell$. For $i=1,\ldots,K$, let $\calC_i$ denote the following set of constraints.
	\begin{align*}
	\sum_{f \in \calF_{\{i,U\}}}y_{\{i,f\}}(U) &= \sum_{\ell \in \calI_U} x_{U}^{(i)}(\ell), ~~ \text{for } U \in \cup_{\ell=1}^\beta \calU_\ell^{(i)},&\nonumber\\
	\sum_{U \in \calU_{\{i,f \}}} y_{\{i,f\}}(U)&= r,~~ \text{for }  f \in \Omega^{(i)},&\nonumber\\
	x_U^{(i)}(\ell), \ y_{\{i,f \}}(U) &\geq 0, ~~ \text{for } U \in \calU_{\ell}^{(i)}, \ \ell \in [\beta], \ f\in \Omega^{(i)}. & \nonumber
	\end{align*}
	Then, the original LP can be compactly rewritten as
	\begin{align}
	\label{eq:LP_MCF2}
	\min ~~& \sum_{\ell=1}^{\beta} \sum_{U \in \calU_\ell} x_{U}(\ell) & \\	
	 \text{s.t. }~~  x_{U}^{(i)}(\ell) &\leq x_{U}(\ell)  ~~ \text{for } U \in \calU_\ell^{(i)}, \ \ell \in [\beta], i \in [K], &\nonumber\\
	\sum_{U \in \calU_\ell} x_U(\ell) &\leq |\Pi_\ell|, ~~ \text{for } \ell \in [\beta],& \nonumber\\
	& \calC_1, \calC_2, \dots, \calC_K.& \nonumber
	\end{align}
%
It is not too hard to see that the LPs in  (\ref{eq:simplfied_linear_program})  and (\ref{eq:LP_MCF2}) are equivalent. The only difference with (\ref{eq:simplfied_linear_program}) is the introduction of variables $x^{(i)}_U(\ell)$ (for appropriate ranges of $i$, $U$ and $\ell$) such that the second set of inequality constraints in (\ref{eq:simplfied_linear_program}) are replaced by equality constraints. Moreover, the original constraints are maintained by setting $x_U^{(i)}(\ell) \leq x_U(\ell)$. 

	By the Slater's constraint qualification condition \cite{boyd2004convex}, we know that if the primal LP is feasible, then strong duality holds and the primal and dual optimal values are the same. Thus, we proceed by considering the dual of the LP in (\ref{eq:LP_MCF2}) with respect to the constraints that involve the variables $x_U(\ell)$. The Lagrangian $\mathcal{L}(\{x_U(\ell),\ x^{(i)}_U(\ell),\ \lambda_U^{(i)}(\ell)\}_{i \in U, U \in \calU_\ell, \ell \in [\beta]},\{\zeta_\ell\}_{\ell \in [\beta]})$ can be expressed as
	\begin{align}
	\mathcal{L} &\triangleq \sum_{\ell=1}^{\beta} \sum_{U \in \calU_\ell} x_{U}(\ell)
	+ \sum_{\ell=1}^{\beta} \sum_{U \in \calU_\ell} \sum_{i \in U} \lambda_U^{(i)}(\ell) \left(x_{U}^{(i)}(\ell) - x_{U}(\ell) \right) \nonumber\\
	&+ \sum_{\ell=1}^{\beta}  \zeta_\ell \left(\sum_{U \in \calU_\ell}x_{U}(\ell) - |\Pi_\ell|   \right)\nonumber
	\end{align}
	where
	$\lambda_U^{(i)}(\ell)$'s and $\zeta_\ell$'s are nonnegative dual variables.
	It turns out that minimizing the Lagrangian for fixed dual variables can be simplified by defining $\gamma_{U}^{(i)}(\ell) = \lambda_{U}^{(i)}(\ell) / (1+\zeta_\ell)$ for $i \in U$, $U \in \calU_\ell$, and $\ell \in [\beta]$. We define $\Gamma^{(i)} = \{\gamma_U^{(i)}(\ell), \ \ell \in \calI_U, \ U \in \calU_\ell^{(i)} \}$, $\mathbf{x} = \{x_U(\ell), \ U \in \calU_{\ell}, \ \ell \in [\beta] \}$, and $\mathbf{x}^{(i)} = \{x_U^{(i)}(\ell), \ \ell \in \calI_U, \ U \in \calU_\ell^{(i)} \}$.
	The dual function $g(\Gamma^{(1)}, \ldots, \Gamma^{(K)}, \{\zeta_\ell\}_{\ell \in [\beta]})$ is obtained by solving for
	\begin{align*}
	\min_{\mathbf{x}, \mathbf{x}^{(1)}, \ldots, \mathbf{x}^{(K)}} \  &\mathcal{L}\\  \qquad \text{s.t. } ~~~ &\calC_1,\calC_2,\ldots, \calC_K.
	\end{align*}
\noindent It is evident that the dual function $g(\Gamma^{(1)}, \ldots, \Gamma^{(K)}, \{\zeta_\ell\}_{\ell \in [\beta]})$ takes a nontrivial value only if
	\begin{align*}
	\sum_{i \in U} \gamma_U^{(i)}(\ell) = 1, \qquad \forall \ U \in \calU_\ell, \ \ell \in [\beta].
	\end{align*}
	
	The evaluation of $g(\Gamma^{(1)}, \ldots, \Gamma^{(K)}, \{\zeta_\ell\}_{\ell \in [\beta]})$ at a fixed set of dual variables $\Gamma^{(i)}$'s and $\zeta_\ell$'s can therefore be written as
	\begin{align}
		\label{eq:DualSimplified}
		\min_{\mathbf{x}^{(1)}, \ldots, \mathbf{x}^{(K)}} \ & \sum_{\ell=1}^{\beta} \sum_{U \in \calU_\ell} \sum_{i \in U} (1+\zeta_\ell)\gamma_U^{(i)}(\ell) x_{U}^{(i)}(\ell)  - \sum_{\ell=1}^\beta \zeta_\ell |\Pi_\ell| \nonumber \\
		\text{s.t. }\ & \calC_1,\calC_2,\ldots, \calC_K.
	\end{align}
We emphasize that (\ref{eq:DualSimplified}) is still a convex problem and that $\gamma_{U}^{(i)}(\ell) , \zeta_\ell \geq 0$. Let $h_i(\Gamma_i, \{\zeta_\ell\}_{\ell \in [\beta]})$, $i\in [K]$ be
	\begin{align}
	\label{eq:Hlambda}
	h_i(\Gamma_i, \{\zeta_\ell\}_{\ell \in [\beta]})\triangleq &
	\min_{\mathbf{x}^{(i)}} ~~~\sum_{\ell=1}^{\beta} \sum_{U \in \calU^{(i)}_\ell} (1+\zeta_\ell)\gamma_U^{(i)}(\ell)  x_{U}^{(i)}(\ell),\nonumber  \\
	&\text{s.t.  }~~\calC_i .
	\end{align}
 	Then, the dual function becomes
	\begin{align}
	\label{eq:DualLP}
g(\Gamma^{(1)}, \ldots, \Gamma^{(K)},& \{\zeta_\ell\}_{\ell \in [\beta]})  = \\
 &\sum_{i=1}^K h_i(\Gamma_i, \{\zeta_\ell\}_{\ell \in [\beta]}) - \sum_{\ell=1}^{\beta} \zeta_\ell |\Pi_{\ell}|,\nonumber
	\end{align}
	if $\sum_{i \in U} \gamma_U^{(i)}(\ell) = 1$ for all $U \in \calU_\ell, \ \ell \in [\beta]$. 
	We present an approach to maximize the dual function in (\ref{eq:DualLP}) shortly. 
%

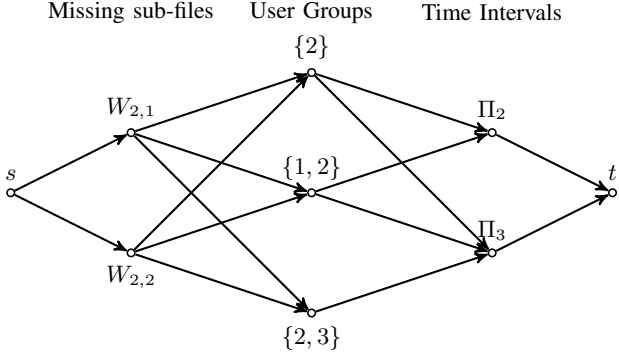
\begin{figure}[t]
	\centering
	\usetikzlibrary{arrows,topaths}
	\begin{tikzpicture}[>=stealth',semithick,auto, scale=0.8]
	\tikzstyle{vtx}  = [circle, minimum width=1pt, draw, inner sep=1pt]

	\tikzstyle{every label}=[font=\bfseries]
	\small{
		\node[vtx,  label=above:{$s$}] 		(s) at (-2,0) {};
		\node[] at (0,3) {Missing sub-files};
		\node[vtx,  label=above:{$W_{2,1}$}] 		(w2) at (0,1) {};
		\node[vtx,  label=below:{$W_{2,2}$}] 		(w3) at (0,-1) {};
		\node[] at (3,3) {User Groups};
		\node[vtx,  label=above:{$\{2\}$}] 					(u1) at (3,2) {};
		\node[vtx,  label=above:{$\{1,2\}$}] 					(u2) at (3,0) {};
		\node[vtx,  label=below:{$\{2,3\}$}]					(u3) at (3,-2) {};
		\node[] at (6,3) {Time Intervals};
		\node[vtx,  label=above:{$\Pi_2$}] 					(l1) at (6,1) {};
		\node[vtx,  label=above:{$\Pi_3$}] 				(l3) at (6,-1) {};
		
		\node[vtx,  label=above:{$t$}] 				(t) at (8,0) {};
		
		\draw[thick,->] 	(s) to node {} (w2);
		\draw[thick,->] 	(s) to node {} (w3);

		\draw[thick,->] 	(w2) to node[label=above:{}] {} (u1);
		\draw[thick,->] 	(w3) to node[label=above:{}] {} (u1);
		\draw[thick,->] 	(w2) to node[label=above:{}] {} (u2);
		\draw[thick,->] 	(w3) to node[label=above:{}] {} (u2);
		\draw[thick,->] 	(w2) to node[label=above:{}] {} (u3);
		\draw[thick,->] 	(w3) to node[label=above:{}] {} (u3);						
		
		\draw[thick,->] 	(u1) to node[label=above:{}] {} (l1);
		\draw[thick,->] 	(u1) to node[label=above:{}] {} (l3);
		\draw[thick,->] 	(u2) to node[label=above:{}] {} (l1);
		\draw[thick,->] 	(u2) to node[label=above:{}] {} (l3);
		\draw[thick,->] 	(u3) to node[label=above:{}] {} (l3);			
		
		\draw[thick,->] 	(l1) to node[label=above:{}] {} (t);
		\draw[thick,->] 	(l3) to node[label=above:{}] {} (t);
		
	}
	\end{tikzpicture}
	\caption{{\small Min-cost flow network associated with subproblem (\ref{eq:Hlambda}) corresponding to the second user, $\calN_2(\Gamma_2, \zeta_2, \zeta_3)$. The constraints and costs are given in the text.}
	\label{Fig:Min_Cost_P_i}}
\vspace{-0.1in}
\end{figure}

	The sub-problem in (\ref{eq:Hlambda}) for fixed $\Gamma_i$ and $\{\zeta_\ell\}_{\ell \in [\beta]}$, is a standard minimum-cost flow problem. The associated flow network corresponding to user $i$, $i\in[K]$, depends on $\Gamma_i$ and $\{\zeta_\ell\}_{\ell \in [\beta]}$ and we denote it by $\calN_i(\Gamma_i, \{\zeta_\ell\}_{\ell \in [\beta]})$. It contains a source node $s$ and three intermediate layers followed by a terminal node $t$ (see Fig. \ref{Fig:Min_Cost_P_i} for an example). The nodes in the first, second, and third layer correspond to missing subfiles in $\Omega^{(i)}$, user groups in $\cup_{\ell\in [\beta]} \calU_\ell^{(i)} $, and time intervals $\{\Pi_\ell: \ \ell \in [\beta]  \text{  and  } i \in U_\ell \}$ respectively. The edges in $\calN_i(\Gamma_i, \{\zeta_\ell\}_{\ell \in [\beta]})$ can be expressed as follows. There are $|\Omega^{(i)}|$ edges going from source node $s$ to each of missing subfiles in $\Omega^{(i)}$. Also, for each $f \in \calF_{\{i,U\}}$ there is an edge going from missing subfile node $f$ to user group node $U$. Furthermore, there is an edge going from user group $U \in \cup_{\ell\in [\beta]} \calU_\ell^{(i)} $ to time interval $\Pi_\ell$ for each $\ell \in \calI_U$. Finally, corresponding to each time interval in $\{\Pi_\ell: \ \ell \in [\beta]  \text{  and  } i \in U_\ell \}$ there is an edge going from this time interval to the terminal node $t$.
	
	In flow network $\calN_i(\Gamma_i, \{\zeta_\ell\}_{\ell \in [\beta]})$, $i \in [K]$, a zero cost is assigned to all edges except those from the user group nodes to the time intervals.
	The cost of the edge between user group $U$ and time interval $\Pi_\ell$ is $(1+\zeta_\ell)\gamma_U^{(i)}(\ell)$. The edge between time interval $\Pi_\ell$ and the terminal node has a capacity constraint of $|\Pi_\ell|$ and the edge between the source node and a missing subfile has a capacity constraint of $r$; the other edges have no capacity constraint. The variable $x_U^{(i)}(\ell)$ is the amount of flow carried by the edge from user group $U$ to time interval $\Pi_\ell$. The source injects a flow of value $|\Omega^{(i)}|r$ which needs to be absorbed in the terminal.

We emphasize that minimum cost network flow algorithms have been subject of much investigation \cite{rk} within the optimization literature and large scale instances can be solved very quickly. For our work, we leverage  Capacity Scaling algorithms within the open-source LEMON package \cite{lemon}. 

\subsubsection{Maximizing the dual function}
	The dual function in (\ref{eq:DualLP}) is concave (as it can be expressed as the pointwise infimum of a family of affine functions of the dual variables \cite{boyd2004convex}). We exploit the projected subgradient method to maximize the dual function iteratively. Let $x_U^{(i)}(\ell, n-1)$ for all $i \in [K], U \in \calU_{\ell}$ denote the optimal point of (\ref{eq:Hlambda}) when solved for $i \in [K]$ at the $n-1$ iteration. Let $\{\gamma_U^{(i)}(\ell, n-1), \ \zeta_\ell(n-1), \  \forall \ U \in \calU^{(i)}_\ell, \ \ell \in [\beta], i \in [K] \}$ denote a dual feasible point of (\ref{eq:DualLP}) at the $(n-1)$-th iteration.

According to the subgradient method, at the $n$-th iteration, for $i \in [K]$, we first compute
	\begin{align*}
	&\tilde{\gamma}_U^{(i)}(\ell, n) = \gamma_U^{(i)}(\ell,n-1) + \theta_{n} x_U^{(i)}(\ell,n)(1+\zeta_\ell(n-1)), \\
	&\tilde{\zeta}_\ell(n) = \zeta_\ell(n-1)+\theta_{n} (\sum_{U \in \calU_\ell} \sum_{i\in U} \gamma_U^{(i)} (\ell, n)x_U^{(i)} (\ell, n) - |\Pi_\ell|),
	\end{align*}
	where $\theta_n$ is the step size. These intermediate variables are projected onto the feasible set  and primal recovery is performed by the method of \cite{sherali1996recovery}. The details can be found in the Appendix \ref{app::primal_recovery}. Numerical results appear in Section \ref{sec:discussion}.


\vspace{-0.1in}	
\section{Online Asynchronous Coded Caching}
\label{sec:online_sec}
In the online scenario, at time $\tau$ only information about the already arrived requests are known to the server, i.e., it only knows $T_i$, $d_i$ and $\Delta_i$ for $i \in [K]$ such that $T_i \leq \tau$.
Ideally, one would want to design an online algorithm that is guaranteed to be feasible whenever the corresponding offline version is feasible. However, this appears to be a hard problem. Specifically, routinely used algorithms such as earliest-deadline-first (EDF) do not have this property \cite{ghasemiThesis}. In the upcoming subsection we demonstrate that the online solution requires additional ideas from a coding standpoint.

\vspace{-0.1in}
\subsection{Necessity of coding across missing subfiles of a user}
\label{sec:necessity_coding}
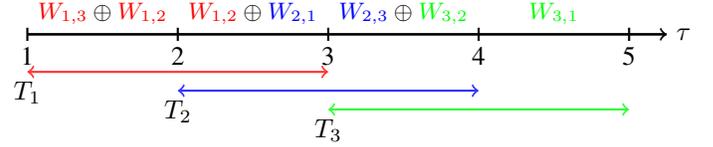
\begin{figure}[t]
		\centering
		\begin{tikzpicture}
		\draw[thick, ->] (2,0) -- (3,0) node[above] {\small{$\textcolor{red}{W_{1,3}}\oplus \textcolor{red}{W_{1,2}}$}} -- (5,0) node[above] {\small{$\textcolor{red}{W_{1,2}}\oplus \textcolor{blue}{W_{2,1}}$}}-- (7,0) node[above] {\small{$\textcolor{blue}{W_{2,3}}\oplus \textcolor{green}{W_{3,2}}$}} -- (9,0) node[above] {\small{$\textcolor{green}{W_{3,1}}$}} --(10.5,0) node[right] {$\tau$};
		\draw [thick,
		postaction={
			draw,
			decoration=ticks,
			segment length=2cm,
			decorate,
		}
		] (2,0) -- (10.5,0);
		\foreach \tick in {1,...,5}
		\node at (2*\tick,0) [below=1pt] {\tick};
		
		\draw[thick, <->, color=red!80] (2,-0.5) node[below] {\textcolor{black}{$T_1$}} -- (6,-0.5);
		\draw[thick, <->, color=blue!80] (4,-0.75) node[below] {\textcolor{black}{$T_2$}}-- (8,-0.75);
		\draw[thick, <->, color=green!80] (6,-1)node[below] {\textcolor{black}{$T_3$}} -- (10,-1);
		\end{tikzpicture}
		\caption{\label{Fig:N_3_K_3_M_1_online} {\small Online solution corresponding to the Example \ref{exmp::coding accros the same file}. Note that the server is forced to transmit $W_{1,2} \oplus W_{1,3}$ at $\tau=1$. }}
		\vspace{-0.2in}
	\end{figure}

	\begin{example}
		\label{exmp::coding accros the same file}
		Consider a system with $N=K=3$ and $M_i=1$ with $Z_i = \{W_{n,i}: \ n \in [N] \}$ for $i \in [K]$ (also depicted in Fig. \ref{Fig:block_coded_caching}).
		The arrival times and deadlines of the users are $T_i=i$, and $\Delta_i = 2$ for $i \in [K]$ (as shown in Fig. \ref{Fig:N_3_K_3_M_1_online}). We assume that user $i$ is interested in files $W_i$ for $i \in [K]$ and that transmitting a subfile takes a single time slot, i.e., $r=1$.

Suppose that the server does not code across any user's missing subfiles. At $\tau=1$, it has the choice to transmit either $W_{1,2}$ or $W_{1,3}$. We emphasize that it has to transmit either of these as the deadline for user 1 is $T_1 + \Delta_1 = 3$ . If the server transmits $W_{1,3}$, then consider the scenario where $(T_3,\Delta_3) = (2,2)$ and $(T_2,\Delta_2)=(3,2)$, i.e., the third user's request comes at $\tau =2$ and the second user's request comes at $\tau = 3$. In this case, the server is forced to transmit $W_{1,2}$ at $\tau=2$, which implies that user 3 misses its deadline. On the other hand, if the server transmits $W_{1,2}$ at $\tau = 1$, then $(T_2,\Delta_2) = (2,2)$ and $(T_3,\Delta_3)=(3,2)$ will cause user 2 to miss its deadline.
%
%
\end{example}

This issue can be circumvented if we transmit a linear combination of both $W_{1,2}$ and $W_{1,3}$ in the first time slot as shown in Fig \ref{Fig:N_3_K_3_M_1_online}. Intuitively, this is the correct strategy since transmitting $W_{1,3} \oplus W_{1,2}$ allows the server to hedge its bets against the identity of the next request arrival. This example demonstrates that coding across missing subfiles of user $1$ is strictly better than the alternative. We emphasize that the synchronized model of \cite{maddahN14} and the offline scenario do not require this.

Accordingly, for the online scenario we treat each missing subfile $W_{d_i, f}$ as an element of a large enough finite field $\mathbb{F}$. This allows us to consider linear combinations of the missing subfiles over $\mathbb{F}$. Note that any equation of the form
\begin{align}
\bigoplus_{i\in U} \bigoplus_{f\in \calF_{\{i,U\}}} \alpha_{\{i,f\}} W_{\{d_i,f\}}, \label{eq:RLC_eq}
\end{align}
where the coefficients $\alpha_{\{i,f\}}$ belong to the field $\mathbb{F}$ and $\bigoplus$ represents $\mathbb{F}$-addition is also an all-but-one equation from which user $i$ can recover $\bigoplus_{f\in \calF_{\{i,U\}}} \alpha_{\{i,f\}} W_{\{d_i,f\}}$. 

\vspace{-0.1in}
\subsection{Recursive LP based algorithm}
\label{sec:online_algorithms}

In the online scenario at time $\tau$ our only decision is to transmit an equation in the time slot $[\tau, \tau + 1)$. In particular, it is possible that a request arrives at $\tau + 1$ and that can change the situation drastically. It makes intuitive sense to transmit equations that benefit a large number of users. However, we also need to take into account the deadline constraints of each user. These requirements need to be balanced. At the top level, our approach can be summarized as follows.
\begin{itemize}[wide, labelwidth=!, labelindent=0pt]
\item {\it Solving a linear program when a user request arrives.} We solve an LP which is similar to (\ref{eq:simplfied_linear_program}) {\it each time} a new user request comes into the system. This specifies a set of $x_U(\ell)$  and $y_{i,f}(U)$ variables. However, in the offline case, the ordering of the $x_U(\ell)$'s within an interval does not matter ({\it cf.} Remark \ref{remark:offline_ordering}). In the online case, this is no longer true. As we have no knowledge of future arrivals, it becomes important to choose the ``best" user group for the time slot in which transmission needs to take place. 

\item {\it Deciding which user group to pick.} Based on the $x_U(\ell)$ variables we first decide a candidate list of feasible user groups that can be chosen for transmission at each time slot. Suppose user group $U$ is a candidate. We calculate a metric for $U$ depending upon (i) the benefit of this equation to the participating users within $U$, and (ii) the stringency of the deadlines of the users in $U$. 
    Our measure of stringency for user $i$ is the ratio of the remaining number of missing packets of user $i$ to the number of remaining time-slots for user $i$. If the calculated metric for $U$ is above a pre-defined threshold then an equation corresponding to $U$ (of the form in (\ref{eq:RLC_eq})) is transmitted.

\item {\it Updating variables and continuing recursively.} Following this, we update certain variables and the process continues for each time slot thereafter. When the next user request arrives into the system, the history of the variable assignments is used to solve a new LP (similar to (\ref{eq:simplfied_linear_program})), and the process continues recursively.

\end{itemize}


\subsubsection{Measuring the benefit of user group $U$ to user $i$}
As we need to commit to a user group at each time instant in the online case, we first discuss how we can measure the benefit to a user $i \in U$ if user group $U$ is chosen for transmission at a given time slot. In particular, if $U$ has been used for transmission in the past, then the current transmission may be less beneficial to some of the users or of no benefit. We demonstrate this by means of the following example.

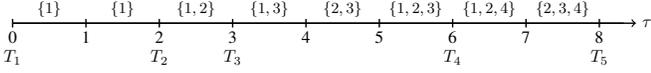
\begin{figure}[t]
	\centering
	\scalebox{0.65}{
		\begin{tikzpicture}
		%
		\draw[thick, ->] (0,0) -- (0.75,0) node[above] {\small{$\{1\}$}} -- (1.5,0) -- (2.25, 0) node[above]  {\small{$\{1\}$}}-- (3,0) -- (3.75, 0) node[above] {\small{$\{1,2\}$}}-- (4.5,0) -- (5.25, 0) node[above] {\small{$\{1,3\}$}} -- (6,0) -- (6.75, 0) node[above] {\small{$\{2,3\}$}} --(7.5,0) -- (8.25, 0) node[above] {\small{$\{1,2,3\}$}} -- (9,0) -- (9.75, 0) node[above] {\small{$\{1,2,4\}$}} --(10.5,0) -- (11.25, 0) node[above] {\small{$\{2,3,4\}$}} -- (12, 0) -- (12.75, 0) node[right] {$\tau$};
		\draw [thick,
		postaction={
			draw,
			decoration=ticks,
			segment length=1.5cm,
			decorate,
		}
		] (0,0) -- (12.5,0);
		\foreach \tick in {0,...,8}
		\node at (1.5*\tick,0) [below=1pt] {\tick};
		
		\node at (0,-0.7) {$T_1$} {};
		\node at (3,-0.7) {$T_2$} {};
		\node at (4.5,-0.7) {$T_3$} {};
		\node at (9,-0.7) {$T_4$} {};
		\node at (12,-0.7) {$T_5$} {};
		\end{tikzpicture}
	}
	\caption{\label{Fig:diffrent_benefit} {\small An illustration of arrival times and user groups associated with the already submitted equations upon time $\tau = 8$ in Example \ref{Exm::different_benefits}. Associated with each user group in each time slot, an equation has been submitted by the server at the same time slot.}}
	\vspace{-0.2in}
\end{figure}

\begin{example}
	\label{Exm::different_benefits}
	Consider a system $N=K=5$, $M_i=2$ for all users $i \in [K]$, and $r=1$. The placement scheme is the same as \cite{maddahN14} so that each file is divided to $F = 10$ subfiles and each user misses $6$ subfiles. The cache content and missing subfiles are specified in Table \ref{tab:placement_MN_KN5}.
	We assume that the current time is $\tau = 8$ and that the request of users $1,\dots,4$ have arrived to the server. More specifically, we have $T_1=0, T_2=2, T_3 = 3, T_4 = 6$ with deadlines $\Delta_i=15$ for all users $i\in [K]$.
The server has already transmitted eight equations  corresponding to the user groups depicted in Fig. \ref{Fig:diffrent_benefit}.

Suppose that the server considers scheduling the user group $\{2,3,5\}$ at $\tau=8$. We note here that users 2 and 3 have already participated in previous transmissions. Thus, it needs to be determined how beneficial this equation is to each user. Considering user 2, as shown in Table \ref{tab:placement_MN_KN5}, it can recover a linear combination of $\{W_{\{d_2,f\}}, \ f = 2,8,9 \}$ (from user group $\{2,3\}$), $W_{\{d_2,2\}}$ (from user group $\{1,2,3\}$) and $W_{\{d_2,8\}}$ (from user group $\{2,3,4\}$) from the prior equations; this in turn implies that it can also recover $W_{\{d_2,9\}}$ by solving linear equations. However, note the the user group $\{2,3,5\}$ also results in user 2 recovering $W_{\{d_2,9\}}$. Thus, from user 2's perspective, this equation is of no benefit. A similar argument shows that this user group does not benefit user 3 either.
\end{example}
 We observe at this point that the modified LP in Section \ref{sec:modified_LP_offline}, can be used here to determine a given user's benefit. In particular, the past history of the transmission contains information on the total time allocated to a given user group $U$.  At time $\tau$, if we consider transmitting user group $\hatU$, we can add it to the set of user groups and update its time allocation. Following this, for each user the LP in (\ref{eq:fixed_u_linear_program}) can be used to measure the number of subfiles that can be transmitted for it, were $\hatU$ to be chosen.


Let $\calU_{\text{sent}}(\tau)$ denote the set of user groups chosen for time $\leq \tau$, $\hatU$ be a user group under consideration at time $\tau$, and let $\tilde{\calU}_{\text{sent}}(\tau) = \calU_{\text{sent}}(\tau) \cup \{\hatU\}$. Let $\tilde{z}_U(\tau)$ denote the time allocated to user group $U \in \tilde{\calU}_{\text{sent}}(\tau)$, where $\tilde{z}_{\hatU}(\tau)$ is incremented by one if $\hatU \in \tilde{\calU}_{\text{sent}}(\tau)$, otherwise it is set to 1. Consider the following LP.
\begin{align}
\label{eq:compute_eta}
\max _{\{\tilde{y}_{\{i,f\}}({U})\}} \sum_{{U} \in \tilde{\calU}_{\text{sent}}(\tau), {U} \ni i} &\sum_{f \in \calF_{\{i,{U}\}}} \tilde{y}_{\{i,f\}}({U}) & \\
\text{s.t. }~\sum_{f\in \calF_{\{i,{U}\}}} \tilde{y}_{\{i,f\}}({U}) & \leq \tilde{z}_{U}(\tau)~ \text{for } \ {U} \in \tilde{\calU}_{\text{sent}}(\tau), U \ni i &\nonumber\\
\sum_{{U} \in \calU_{\{ i, f\}} \cap \tilde{\calU}_{\text{sent}}(\tau)}  \tilde{y}_{\{i,f\}}({U}) &\leq r  ~~  \text{for } f\in \Omega^{(i)},&\nonumber\\
\tilde{y}_{\{i,f\}}({U})& \geq 0.&\nonumber
\end{align}
Now suppose that we have already tracked the number of useful packets for user $i$ until time $\tau$. Then the above LP can be used to determine the benefit of transmitting user group $\hatU$ at time $\tau$.

\begin{table}[!t]
	\begin{center}
		\begin{tabular}{|c | c | c|}
			\hline
			User & Cache content indices & Missing subfiles indices\\
			\hline
			$1$ & $1,2,3,4$ & $5,6,7,8,9,10$ \\
			\hline
			$2$ & $1,5,6,7$ & $2,3,4,8,9,10$ \\
			\hline
			$3$ & $ 2,5,8,9$ & $1,3,4,6,7,10$ \\
			\hline
			$4$ & $3,6,8,10$ & $1,2,4,5,7,9$ \\
			\hline
			$5$ & $4,7,9,10$ & $1,2,3,5,6,8$ \\
			\hline
		\end{tabular}
	\end{center}
	\caption{\label{tab:placement_MN_KN5} Cache content and missing subfiles of Example \ref{Exm::different_benefits}. Cache contents $Z_i = \{ W_{\{n, f\}},\ n \in [5], f\in \text{ second column} \}$ and missing subfiles $\{W_{\{d_i,f\}}, f\in \text{ third column} \}$.}
\end{table}

\begin{figure}[t]
	\centering
	\usetikzlibrary{arrows,topaths}
	\begin{tikzpicture}[scale=0.75]
	\tikzstyle{vtx}  = [circle, minimum width=1pt, draw, inner sep=1pt]

	\tikzstyle{every label}=[font=\bfseries]
	\small{
		\node[vtx,  label=above:{$s$}] 		(s) at (-2,0) {};
		\node[] at (0,2.5) {Subfiles in $\Omega^{(i)}$};
		\node[vtx,  label=above:{$W_{d_i,f_1}$}] 		(w2) at (0,1.5) {};
		\node[vtx,  label=below:{$W_{d_i,f_l}$}] 		(w3) at (0,-1.5) {};
		\node[] at (4,2.5) {User Groups in $\tilde{\calU}_{\text{sent}}(\tau)$};
		\node[vtx,  label=above:{$U$}] 					(u1) at (4,1.5) {};
		\node[vtx,  label=above:{$U'$}] 					(u2) at (4,0) {};
		\node[vtx,  label=below:{$U''$}]					(u3) at (4,-1.5) {};
		%
		\node[vtx,  label=above:{$t$}] 				(t) at (6,0) {};
		
		\draw[thick,->] 	(s) to node {} (w2);
		\draw[thick,->] 	(s) to node {} (w3);

		\draw[thick,->] 	(w2) to node[label=above:{\small $\tilde{y}_{\{i,f_1\}}(U)$}] {} (u1);
		\draw[thick,->] 	(w3) to node[label=above:{}] {} (u1);
		\draw[thick,->] 	(w2) to node[label=above:{}] {} (u2);
		\draw[thick,->] 	(w3) to node[label=above:{}] {} (u2);
		\draw[thick,->] 	(w2) to node[label=above:{}] {} (u3);
		\draw[thick,->] 	(w3) to node[label=below:{$\tilde{y}_{\{i,f_l\}}(U'')$}] {} (u3);						
		
		
		\draw[thick,->] 	(u1) to node[label=above:{}] {} (t);
		\draw[thick,->] 	(u2) to node[label=above:{}] {} (t);
		\draw[thick,->] 	(u3) to node[label=above:{}] {} (t);
		
		\node[] at (0,0) {\vdots};
		\node[] at (4,1.1) {\vdots};
		\node[] at (4,-0.7) {\vdots};
		
	}
	\end{tikzpicture}
	\caption{{\small Max flow network associated with LP in (\ref{eq:compute_eta}).}}
	\label{Fig:max_flow_w}
	\vspace{-0.1in}
\end{figure}
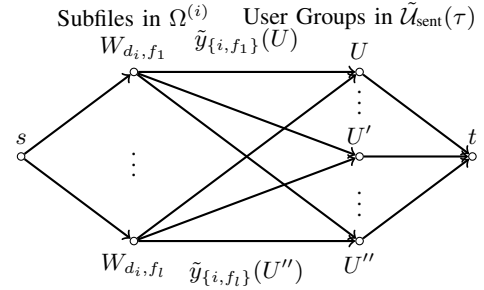

\begin{remark}
	\label{rem::integrality_of_ys}
	The LP in (\ref{eq:compute_eta}) can also be expressed as a maximum flow problem. The associated flow network consists of a source node $s$, a node for each $f \in \Omega^{(i)}$, a node for each user group $U \in \tilde{\calU}_{\text{sent}}(\tau)$, and a terminal node $t$. There are edges with capacity $r$ going from $s$ to each $f \in \Omega^{(i)}$ and edges from $f\in \Omega^{(i)}$ to node $U \in \tilde{\calU}_{\text{sent}}(\tau)$ if $f \in \calF_{\{i,{U}\}}$. The flow on such an edge is $\tilde{y}_{\{i,f\}}(U)$. Moreover, from each node $U \in \tilde{\calU}_{\text{sent}}(\tau)$ to $t$ there exist an edge of capacity $\tilde{z}_U(\tau)$. These capacity constraints model the first two inequality constraints in (\ref{eq:compute_eta}). Fig. \ref{Fig:max_flow_w} illustrates an example of this network. It is well-known that if all capacities in a flow network are integers, there exists an integral maximum flow (\cite{kleinberg2006algorithm}, Chapter 7). Therefore, there exists an integral solution for $\tilde{y}_{\{i,f\}}(U)$'s in (\ref{eq:compute_eta}) if $\tilde{z}_U(\tau)$'s are integers.
\end{remark}

\subsubsection{Solving LP upon user arrival}
Consider a time  $\tau=T_k$ when the request of the $k$-th user arrives at the server.  We let $\calU_{\text{sent}}(\tau)$ be the set of user groups associated with the previously transmitted equations. We also let $z_U(\tau)$ be the total time allocated to equations corresponding to user group $U$ prior to time $\tau$. Thus, if in time interval $[\tau, \tau+1 )$ the server transmits an equation that exclusively benefit users in $U$ then $z_U(\tau+1) = z_U(\tau) + 1$ otherwise $z_U(\tau + 1) = z_U(\tau)$.
Time intervals $\Pi_{1, k}, \ldots, \Pi_{\beta_k, k}$ are formed by the set of times in
\beqno
\left\{T_k \right\} \cup \left\{T_i+\Delta_i: i\in [k],~ T_i+\Delta_i > T_k\right\}.
\eeqno
As in the offline case in (\ref{eq:simplfied_linear_program}), the sets of active users $U_{\ell,k}$, user groups $\calU_{\ell, k}$ and $\calI_U^{(k)}$ are defined corresponding to these time intervals, e.g., $U_{\ell, k}$ is the set of active users in $\Pi_{\ell, k}$. Moreover, $\calV_k$ is a set of user groups that either already have been transmitted or might be transmitted after $\tau = T_k$. That is $\calV_k = \calU_{\text{sent}}(\tau) \cup \{\calU_{\ell, k}:~\ell \in [\beta_{k}] \}$. The variables $x_U(\ell)$'s and $y_{\{i,f\}}(U)$'s have the same interpretation as the offline case.
With these variables, the server solves the following LP.
\begin{align}
\label{eq:LP_rec}
\min_{\{x_U(\ell),~y_{\{i,f\}}(U)\}}  & ~\sum_{\ell = 1}^{\beta_{k}} \sum_{U \in \calU_{\ell, k}} x_{U}(\ell)&  \\
\text{ s.t. }~\sum_{U \in \calU_{\ell,k}} x_U(\ell) &\leq  |\Pi_{\ell,k}|,~~ \text{for} \ \ell = 1, \ldots,\beta_{k} &\nonumber \\
\sum_{f \in \calF_{\{i,U\}}}y_{\{i,f\}}(U) &\leq \sum_{\ell \in \calI^{(k)}_U} x_{U}(\ell) + z_U(T_k) ~ \text{for} \  i \in U,  U \in \calV_k, &\nonumber\\
\sum_{U \in \calU^{}_{\{i,f \}}} y_{\{i,f\}}(U)&= r, ~~ \text{for} \ f \in \Omega^{(i)},\ \forall \ i \in \cup_{\ell=1}^{\beta_{k}} U_{\ell,k},&  \nonumber\\
x_U(\ell) ,~ y_{\{i,f \}}(U)&\geq  0,  \text{ for } i \in [k], \ell \in [\beta], U \in \calV_k.&\nonumber
\end{align}
%
An important feature of time intervals $\Pi_{1,k}$, \ldots, $\Pi_{\beta_k,k}$ is that these time intervals end at a deadline and except the first time interval $\Pi_{1,k}$ that starts with arrival time $T_k$, the other time intervals start with a deadline. Thus, we have $U_{\ell+1, k} \subset U_{\ell,k}$, i.e., the set of active users in interval $\Pi_{\ell+1,k}$ is a subset of the active users in interval $\Pi_{\ell,k}$ for the range of $\ell$.

Next, the server creates a list of candidate user groups. Let $\{x^*_U(\ell), \ \forall \ U \in \calU_\ell, \ \ell = 1,\ldots, \beta_k \}$ be the solution of (\ref{eq:LP_rec}) and let $\calX^* = \{x^*_U(\ell): \ x^*_U(\ell) \geq 1 \}$. 
The elements of $\calX^*$ are first ordered based on time intervals. Then, among the elements with the same time interval, they are ordered based on length of user group. Therefore, for two elements $x^*_{U}(\ell), x^*_{U'}(\ell') \in \calX^*$ we say $x^*_U(\ell)$ is before $x^*_{U'}(\ell')$ if $\ell < \ell'$, or if $\ell = \ell'$ and $|U| \geq |U'|$. We let $\calX_{\text{sorted}}^*$ denote the sorted version of $\calX^*$ using this procedure. Let $v_i(\tau)$ be the number of missing packets (subfiles when $r=1$) that have been transmitted for user $i$ until time $\tau$; this value is tracked in Algorithm \ref{Alg:Onlin_LP_rec}.

Note that user $i$ needs to recover $r|\Omega^{(i)}| - v_i(\tau)$ missing packets and it has $T_i + \Delta_i - \tau$ time slots to obtain them. We use the ratio of these quantities as a measure of the stringency of the deadline of user $i$. Let $w_{\{i,\hat{U}\}}$ denote the total number of missing packets of user $i$ that can be communicated by the user groups chosen thus far and by picking $\hatU$ at time $\tau$. The LP in (\ref{eq:compute_eta}) allows us to compute $w_{\{i,\hat{U}\}}$ and in turn $w_{\{i,\hat{U}\}}(\tau) - v_i(\tau)$.
Therefore the metric $\eta_U(\tau)$ is obtained by the following weighted sum.
\begin{align*}
\eta_U(\tau) \triangleq\sum_{i \in U}  \frac{\left(r|\Omega^{(i)}| - v_i(\tau)\right)}{T_i+\Delta_i - \tau}(w_{\{i,U\}}(\tau) - v_i(\tau)).
\end{align*}
At time $\tau = T_k$, the server picks the first element $x^*_U(\ell) \in \calX_{\text{sorted}}^*$ such that $\eta_U(\tau) \geq \eta_0$ for some threshold $\eta_0$ and transmits an equation corresponding to it.
Unlike the synchronous case, we choose a random linear combination of all missing packets of user $i$ that can be transmitted by user group $U$.

When $r > 1$, we subdivide a missing subfile into $r$ packets that are denoted $W_{\{d_i,f,j\}}$ for $j = 1, \dots, r$.
Thus, the server transmits
\begin{align*}
\bigoplus_{i \in U} \bigoplus_{f \in \calF_{\{i,U\}}} \bigoplus_{j=1}^r \alpha_{\{i,f,j, m\}} W_{\{d_i,f,j\}},
\end{align*}
at time interval $[\tau, \tau + 1)$ where $m$ denotes the $m$-th equation transmitted by the server and $\alpha_{\{i,f,j, m\}}$ are chosen independently and uniformly at random from the finite field $\mathbb{F}$. If none of the elements in $x^*_U(\ell) \in \calX_{\text{sorted}}^*$ satisfy $\eta_U(\ell) \geq \eta_0$ then nothing will be transmitted at this time interval.

If a new user request does not come at time $\tau + 1$, then the server updates the user group values and then solves (\ref{eq:compute_eta}) again to decide the user group for the time slot $[\tau + 1, \tau + 2 )$. The process continues this way until the next user request comes when the LP in (\ref{eq:LP_rec}) is solved. The complete details are provided in Algorithm \ref{Alg:Onlin_LP_rec}.

\begin{algorithm}[t]
	\caption{Recursive LP Algorithm}\label{Alg:Onlin_LP_rec}
	\algrenewcommand\algorithmicrequire{\textbf{Input:}}
	\algrenewcommand\algorithmicensure{\textbf{Output:}}
	\begin{algorithmic}[1]
		\Require Caches $Z_i$ for $i \in [K]$,  $\eta_{0}$, $\{T_i, \Delta_i\}$, for $i\in [K]$.
		\State \textbf{Initialization}:
		\State set $\calU_{\text{sent}}(0) \gets \emptyset$, $\calX_{\text{off}}\gets \emptyset$, $\ell_{\text{off}} \gets 0$, $m\gets 1$ and $k\gets 1$.
		\State set $\calM_i = \emptyset$, and $v_i(0) = 0$ for $i=1,\ldots,K$.
		\For{$\tau= 0,1, 2,  \ldots,  T_{\max}$}
		\If{$\tau = T_i+\Delta_i$ and $v_i(\tau) < r|\Omega^{(i)}|$ for some $i$}
		\State return {\ttfamily INFEASIBLE}.
		\EndIf
		\If{ $\tau = T_i$ (a new user makes request) for some $i$}
		\State $k = \arg\max_{i \in [K]} \tau = T_i$
		\State Solve LP (\ref{eq:LP_rec}). Form $\calX^*$ and then $\calX^*_{\text{sorted}}$.
		\EndIf
		\State If $\tau = T_i$ or $\tau = T_i+\Delta_i$ for some $i$ then $\ell_{\text{off}} \gets \ell_{\text{off}} + 1$.
		\If{$\calX^*_{\text{sorted}} \neq \emptyset$}
		\State \hspace{-5mm} Pick first in order $x^*_{U^*}(\ell) \in \calX^*_{\text{sorted}}$ with $\eta_{U^*}(\tau) \geq \eta_{0}$.
		\State \hspace{-5mm}Randomly select $\alpha_{\{i, f, j, m\}}$'s from $\mathbb{F}$ and send
		\begin{align*}
        \bigoplus_{i \in U} \bigoplus_{f \in \calF_{\{i,U\}}} \bigoplus_{j=1}^r \alpha_{\{i,f,j, m\}} W_{\{d_i,f,j\}},
		\end{align*}

		\State \hspace{-5mm} If ${U^*} \in \calU_{\text{sent}}(\tau)$ then $z_{U^*}(\tau + 1)\gets z_{U^*}(\tau)+1$, otherwise $z_{U^*}(\tau + 1)=1$ and $\calU_{\text{sent}}(\tau + 1) \gets \calU_{\text{sent}}(\tau) \cup \{{U^*}\}$
		
		\State \hspace{-5mm}If $\tilde{x}_{U^*}(\ell_{\text{off}}) \in \calX_{\text{off}}$ then $\tilde{x}_{U^*}(\ell_{\text{off}}) \gets \tilde{x}_{U^*}(\ell_{\text{off}}) + 1$, otherwise $\tilde{x}_{U^*}(\ell_{\text{off}})=1$ and $\calX_{\text{off}} \gets \calX_{\text{off}} \cup \{\tilde{x}_{U^*}(\ell_{\text{off}})\}$
		\State \hspace{-5mm}Set $x^*_{U^*}(\ell) \gets x^*_{U^*}(\ell) - 1$, if $x^*_{U^*}(\ell) < 1$ remove it from $\calX^*_{\text{sorted}}$.
		\State \hspace{-5mm}For all $i\in {U^*}$, set $v_i(\tau+1) \gets w_{\{i,{U^*}\}}(\tau)$, set $\calM_i \gets \calM_i \cup \{m\}$, then $m \gets m+1$
		\State \hspace{-5mm}for all ${U} \in \calU_{\text{sent}}(\tau) \setminus \{U^*\}$ set $z_{{U}} (\tau + 1) \gets z_{{U}} (\tau) $
		\State \hspace{-5mm}for all $i \in [K]\setminus {U^*}$ set $v_i(\tau + 1) \gets v_i(\tau)$.
		\EndIf
		\EndFor
	\end{algorithmic}
\end{algorithm}

In general, there is no guarantee that Algorithm \ref{Alg:Onlin_LP_rec} will return a feasible schedule if the corresponding offline schedule is feasible. In that sense, Algorithm \ref{Alg:Onlin_LP_rec} can be viewed as a heuristic with good experimental performance. However, if Algorithm \ref{Alg:Onlin_LP_rec} does not return ``{\ttfamily INFEASIBLE}", we can show that a feasible solution for the corresponding offline LP can be identified. 
This fact coupled with usage of the Schwartz-Zippel Lemma allows us to conclude that our algorithm works with high probability if it does not return ``{\ttfamily INFEASIBLE}". The proofs of the following claim and lemma appear in Appendix \ref{proof_claim:online_feasibility} and \ref{proof_lemma:zippleSchwart} respectively.
\begin{claim}
	\label{claim:online_feasibility}
	For user requests, $\{T_i, \Delta_i, d_i\}$, where $i \in [K]$, if Algorithm \ref{Alg:Onlin_LP_rec} does not return ``{\ttfamily INFEASIBLE}" then there exists a feasible integral solution for the offline LP in (\ref{eq:simplfied_linear_program}).
\end{claim}
The following lemma shows that if Algorithm \ref{Alg:Onlin_LP_rec} does not return ``{\ttfamily INFEASIBLE}" then with high probability each user recovers all its missing subfiles from the transmitted equations.
%
%
%
\begin{lemma}
	\label{lemma:zippleSchwart}
	If Algorithm \ref{Alg:Onlin_LP_rec} does not return ``{\ttfamily INFEASIBLE}" then with probability at least $\left(1 - \frac{1}{|\mathbb{F}|}\right)^{r KF}$ all requests will be satisfied within their deadline.
\end{lemma}

Thus, by choosing the field size $|\mathbb{F}|$ large enough, we can make the probability of success as large as we want. We point out that increasing the field size results in a corresponding increase in the computational requirements at the server and the user nodes.
\newcolumntype{P}[1]{>{\centering\arraybackslash}p{#1}}
\begin{table}[t]
	\centering
	\begin{tabular}{|c|P{1.4cm}|P{1.4cm}|P{1.59cm}|P{1.59cm}|}
		\hline
		$(K,t)$ & No. of nodes & No. of edges & Exec. time (min) & Exec. time Orig. (min) \\ \hline
		$(100,2)$ & $986,161$ & $17,643,986$ & $8,026$ & ---\\ \hline
		$(20,4)$ & $178,542$ & $1,778,703$ & $1065$ & ---\\ \hline
		$(40,2)$ & $61,959$ & $567,780$ & $63$ & --- \\ \hline
		$(20,2)$ & $7,542$ & $43,507$ & $1.9$ & $21.9$\\ \hline
		$(10,4)$ & $3,917$ & $29,369$ & $0.8$ & $5.33$ \\ \hline
		$(10,2)$ & $915$ & $3,866$ & $0.08$ & $0.03$\\ \hline
	\end{tabular}
	\caption{\label{tab:ProblemSizes}  {\small Execution time for solving the LP using our approach; we run $1000$ iterations of subgradient ascent. Columns $2$ \& $3$ indicate the size of the associated flow network. The table is ordered by the number of nodes in the flow network.}\vspace{-3mm}}
\end{table}

\vspace{-0.1in}
\section{Simulation Results and Comparisons with prior work}
\label{sec:discussion}
In this section we present simulation results for both the proposed offline and the online algorithms (software are available in \cite{asynch_sotware}).
Prior work in this area is primarily the work of \cite{niesen2015coded} that presents heuristics for the online scenario. However, we note that \cite{niesen2015coded} works with deadlines for subfiles and does not take into account the time required to transmit a packet. It uses intuitively plausible rules to decide the equations transmitted by the server depending on the deadlines of the users.

For both scenarios, the request arrival times $\{T_i, \ i\in [K] \}$ are generated according to a Poisson process with parameter $\lambda F$. 
The arrival time is quantized to the nearest time slot. The deadlines  $\Delta_i$, $i\in[K]$ are generated uniformly at random from the range $[\Delta_{\min}, \Delta_{\max}]$ (these values will be specified for each setting below).

\vspace{-0.1in}
\subsection{Offline scenario simulation}

\begin{figure}[!t]
	\centering
	\begin{tikzpicture}
	\begin{axis}[
	ylabel = \small{Primal Solution},
	xlabel = \small{Iterations},
	ylabel near ticks,
	scale= 0.9,
	xmin = 0,
	xmax = 500,
	ymin = 1900,
	ymax = 3350,
	grid,
	]
	\addplot[line width=0.05cm, color = black] table[x=Itr,y=RateP]{out_K20_t2.txt} ;
	\addlegendentry{Primal value over iters.};
	\addplot[line width=0.04cm, dashed, color = black] table[x=Itr,y=Rc]{out_K20_t2.txt};
	\addlegendentry{Optimal value};
	\end{axis}
	\end{tikzpicture}
	\caption{\label{Fig:Offline_dual_Results_K20_t2}{\small Convergence of primal recovery to the optimal solution for a system with $N=K=20$, $r=1$, and $t=2$. Dashed line is the optimal value obtained by solving (\ref{eq:simplfied_linear_program}).}}
\vspace{-0.1in}
\end{figure}
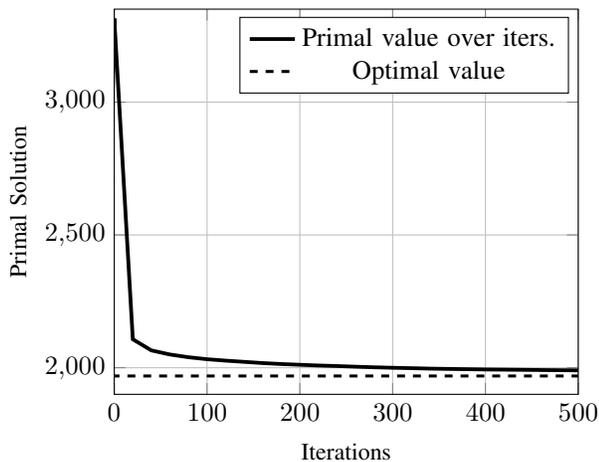

In the first set of simulations we examine the execution time of our approach for various values of $(K,t)$ where $t=KM/N$ is an integer; the placement scheme in  \cite{maddahN14} was used. In these simulations we set $r=1$, $\lambda = 0.4$, $F= \binom{K}{t}$, $\Delta_{\min} = \binom{K-1}{t}$, and $\Delta_{\max} = \binom{K}{t+1}$. 
Table \ref{tab:ProblemSizes} shows the details of the overall execution time and the size of the corresponding flow networks for the various instances. The last column of the table corresponds to the execution time (in MATLAB) of the LP in (\ref{eq:simplfied_linear_program}), while the second-last column corresponds to the execution time of the proposed approach above. It is evident that the proposed approach is significantly faster. In fact, memory requirements make it infeasible to even formulate the problems corresponding to the first three rows in MATLAB.
Fig. \ref{Fig:Offline_dual_Results_K20_t2} shows the convergence of the primal recovery procedure to the actual rate for a system with $N=K=20$, $t=2$, and $r=1$. It can be observed that there is a clear convergence of the solution to the optimal value.

\vspace{-0.1in}
\subsection{Online scenario simulation}
For the online scenario we consider both centralized \cite{maddahN14} and decentralized \cite{maddah2015decentralized} placement schemes for a system with $N=K=6$ and $M=2$ with $\Delta_{\min} = (KM/N)F$ and $\Delta_{\max} = KF$. For each experiment we run $200$ trials for generating the arrivals. For the centralized case, we use the placement scheme of \cite{maddahN14} and the placement is fixed during each experiment. In the decentralized scheme, at each trial the cache content of each user is independently and uniformly chosen as well.

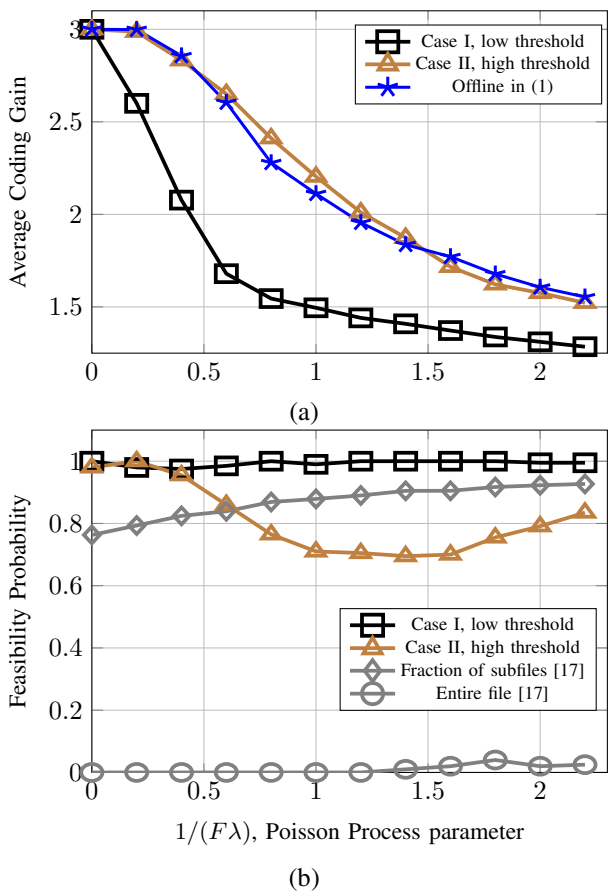
\begin{figure}[!t]
	\centering
	\begin{tabular}{c }
		\begin{tikzpicture}
		\begin{axis}[
		legend style={at={(0.75,1.2)},anchor=north,nodes={scale=0.7, transform shape}},
		ylabel = {\small Average Coding Gain},
		xlabel style={below},
		ylabel style={above},
		ylabel near ticks,
		yscale= 0.8,
		xmin = 0.0,
		xmax = 2.3,
		ymin = 1.25,
		ymax = 3.1,
		grid,
		]
		\addplot[line width=0.05cm, color = black, mark=square, mark size = 4] table[x=lam,y=onGain]{Journal_Cent_K6_t2_low_Mar13.txt} ;
		\addlegendentry{Case I, low threshold};
		\addplot[line width=0.05cm, color = brown, mark=triangle, mark size = 4] table[x=lam,y=onGain]{Journal_Cent_K6_t2_high_Mar13.txt} ;
		\addlegendentry{Case II, high threshold};
		\addplot[line width=0.04cm, color = blue, mark = star, mark size =4] table[x=lam,y=offGain]{Journal_Cent_K6_t2_low_Mar13.txt} ;
		\addlegendentry{Offline in (\ref{eq:simplfied_linear_program})};
		\end{axis}
		\end{tikzpicture}
		\\ (a) \\
		\begin{tikzpicture}
		\begin{axis}[
		legend style={at={(0.73, 0.6)},anchor=north, nodes={scale=0.7, transform shape}},
		xlabel = {\small $1/(F\lambda)$, Poisson Process parameter},
		ylabel = {\small Feasibility Probability},
		xlabel style={below},
		ylabel style={above},
		ylabel near ticks,
		yscale= 0.8,
		xmin = 0.0,
		xmax = 2.3,
		ymin = 0.0,
		ymax = 1.1,
		grid,
		]
		\addplot[line width=0.05cm, color = black, mark = square, mark size = 4] table[x=lam,y=onPro]{Journal_Cent_K6_t2_low_Mar13.txt} ;
		\addlegendentry{Case I, low threshold};
		\addplot[line width=0.05cm, color = brown, mark = triangle, mark size = 4] table[x=lam,y=onPro]{Journal_Cent_K6_t2_high_Mar13.txt} ;
		\addlegendentry{Case II, high threshold};
		\addplot[line width=0.05cm, color = gray, mark = diamond, mark size = 4] table[x=lam,y=onNF]{Journal_Cent_K6_t2_low_Mar13.txt} ;
		\addlegendentry{Fraction of subfiles \cite{niesen2015coded}};
		\addplot[line width=0.05cm, color = gray, mark = o, mark size = 4] table[x=lam,y=onNPro]{Journal_Cent_K6_t2_low_Mar13.txt} ;
		\addlegendentry{Entire file \cite{niesen2015coded}};
		\end{axis}
		\end{tikzpicture}
		\\ (b)
	\end{tabular}
	
	\caption{\label{Fig:K8_M2_Center}{\small Centralized Placement in \cite{maddahN14}: (a) average coding gain over all feasible offline problem instances, (b) feasibility probability of the online algorithm conditioned on feasibility of the offline problem. The placement has been fixed for all trials and at each trial a new arrival time and deadline is generated. In this simulation, we set $\eta_{0} = 0.4 - \frac{0.5}{\lambda }$ and $\eta_0 = 0.8 - \frac{0.2}{\lambda}$ in Case I and II respectively.}}
\vspace{-0.1in}
\end{figure}

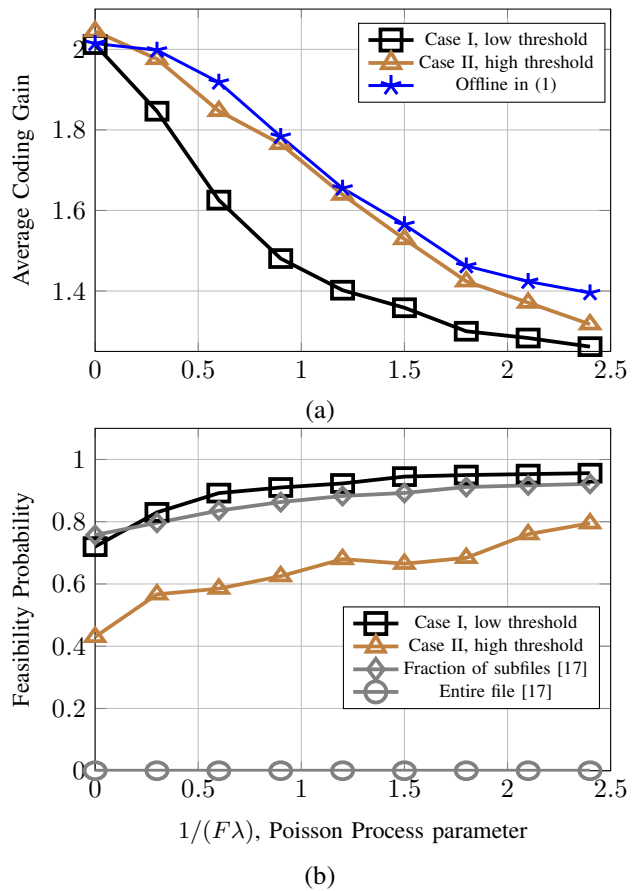
\begin{figure}[!t]
	\centering
	\begin{tabular}{c }
		\begin{tikzpicture}
		\begin{axis}[
		legend style={at={(0.75,1.2)},anchor=north,nodes={scale=0.7, transform shape}},
		ylabel = {\small Average Coding Gain},
		xlabel style={below},
		ylabel style={above},
		ylabel near ticks,
		yscale= 0.8,
		xmin = 0.0,
		xmax = 2.5,
		ymin = 1.25,
		ymax = 2.1,
		grid,
		]
		\addplot[line width=0.05cm, color = black, mark=square, mark size = 4] table[x=lam,y=onGain]{Journal_Dec_K6_t2_low_Mar13.txt} ;
		\addlegendentry{Case I, low threshold};
		\addplot[line width=0.05cm, color = brown, mark=triangle, mark size = 4] table[x=lam,y=onGain]{Journal_Dec_K6_t2_high_Mar13.txt} ;
		\addlegendentry{Case II, high threshold};
		\addplot[line width=0.04cm, color = blue, mark = star, mark size =4] table[x=lam,y=offGain]{Journal_Dec_K6_t2_low_Mar13.txt} ;
		\addlegendentry{Offline in (\ref{eq:simplfied_linear_program})};
		\end{axis}
		\end{tikzpicture}
		\\ (a) \\
		\begin{tikzpicture}
		\begin{axis}[
		legend style={at={(0.73, 0.6)},anchor=north,nodes={scale=0.7, transform shape}},
		xlabel = {\small $1/(F\lambda)$, Poisson Process parameter},
		ylabel = {\small Feasibility Probability},
		xlabel style={below},
		ylabel style={above},
		ylabel near ticks,
		yscale= 0.8,
		xmin = 0.0,
		xmax = 2.5,
		ymin = 0.0,
		ymax = 1.1,
		grid,
		]
		\addplot[line width=0.05cm, color = black, mark = square, mark size = 4] table[x=lam,y=onPro]{Journal_Dec_K6_t2_low_Mar13.txt} ;
		\addlegendentry{Case I, low threshold};
		\addplot[line width=0.05cm, color = brown, mark = triangle, mark size = 4] table[x=lam,y=onPro]{Journal_Dec_K6_t2_high_Mar13.txt} ;
		\addlegendentry{Case II, high threshold};
		\addplot[line width=0.05cm, color = gray, mark = diamond, mark size = 4] table[x=lam,y=onNF]{Journal_Dec_K6_t2_low_Mar13.txt} ;
		\addlegendentry{Fraction of subfiles \cite{niesen2015coded}};
		\addplot[line width=0.05cm, color = gray, mark = o, mark size = 4] table[x=lam,y=onNPro]{Journal_Dec_K6_t2_low_Mar13.txt} ;
		\addlegendentry{Entire file \cite{niesen2015coded}};
		\end{axis}
		\end{tikzpicture}
		\\ (b)
	\end{tabular}
	
	\caption{\label{Fig:K8_M2_Decenter}{\small Decentralized placement scheme for $N=K=6$, $M=2$, and $F=100$: (a) average coding gain over all feasible offline problem instances, (b) feasibility probability of the online algorithm conditioned on feasibility of the offline problem. At each trial cache content of each user is placed randomly and uniformly. In this simulation, we set $\eta_{0} = 0.4 - \frac{0.5}{\lambda }$  and $\eta_{0} = 0.8 - \frac{0.2}{\lambda }$ in Case I and Case II respectively.}}
\vspace{-0.1in}
	
\end{figure}

For each set of generated arrivals, we first run the offline LP to check whether it is feasible. The online algorithm is run only if the offline LP is feasible. The online algorithm requires a threshold $\eta_0$ (see Section \ref{sec:online_algorithms}).  We run simulations with a low threshold (case I) and a high threshold (case II). The coding gain is defined as the ratio of the uncoded rate\footnote{The uncoded rate is simply the total number of missing subfiles of all users normalized by $F$.} to the rate achieved by the system. Fig. \ref{Fig:K8_M2_Center} (a) and Fig. \ref{Fig:K8_M2_Decenter} (a) depict plots of the coding gain vs. $1/(F\lambda)$ in centralized and decentralized cases, respectively. As $\lambda$ decreases, the arrivals are spaced further apart on average, and the coding gain of any scheme is expected to reduce.

The coding gain is computed by taking an average over all instances where a given scheme is feasible. For the offline scheme, this means that we take the average of all instances where it is feasible. For the online algorithm, some of the arrival patterns may result in infeasibility; these instances were not taken into account when computing the average coding gain. This explains why the coding gain of the case II sometimes appears to be higher than the offline algorithm. However, the coding gain of the case I is significantly lower, because of its low threshold.

The feasibility probability of a scheme vs. the arrival rate is plotted in Fig. \ref{Fig:K8_M2_Center} (b) and Fig. \ref{Fig:K8_M2_Decenter} (b) for the centralized and decentralized placement schemes respectively. As expected the low threshold online algorithm has a very high feasibility probability $\approx 1$ for a range of arrival parameters, while the high threshold algorithm has a lower feasibility probability.
Note that the high threshold algorithm (when compared to the low threshold case) only transmits an equation when a large enough number of users benefit from the transmission. Thus, its feasibility probability is lower, but when it is feasible, its coding gain is much higher than the low threshold case.


For both plots, we also include the results of \cite{niesen2015coded}. In this scheme feasibility and coding gain can be traded off by setting a threshold for the defined misfit function (Section III in \cite{niesen2015coded}). We use this scheme by setting the threshold to zero; this is the so-called First-Fit Rule in \cite{niesen2015coded}. The First-Fit rule prefers feasibility over coding gain. The setting in \cite{niesen2015coded} considers a scenario where each subfile has a deadline. We have adapted their algorithm for our case. It can be observed that the feasibility probability of \cite{niesen2015coded} is quite poor. Accordingly, we also plot the fraction of subfiles that meet the deadline; this is somewhat better. The coding gain numbers for \cite{niesen2015coded} are also quite unreliable as the algorithm is infeasible in most cases. Thus, we do not plot it.
\begin{figure}[!t]
	\centering
	\begin{tabular}{c }
		\begin{tikzpicture}
		\begin{axis}[
		legend style={at={(0.82,1.24)},anchor=north,nodes={scale=0.7, transform shape}},
		legend cell align = {left},
		ylabel = {\small Average Coding Gain},
		xlabel style={below},
		ylabel style={above},
		ylabel near ticks,
		yscale= 0.8,
		xmin = 0.0,
		xmax = 2.5,
		ymin = 1.00,
		ymax = 2.0,
		grid,
		]
		\addplot[line width=0.05cm, color = black, mark=square, mark size = 4] table[x=lambda,y=AvgGanOn]{K6N6M2F20lowdate26.txt} ;
		\addlegendentry{Case I};
		\addplot[line width=0.05cm, color = brown, mark=triangle, mark size = 4] table[x=lambda,y=AvgGanOn]{K6N6M2F20highdate30.txt} ;
		\addlegendentry{Case II};
		\addplot[line width=0.04cm, color = blue, mark = star, mark size =4] table[x=lambda,y=AvgGanOff]{K6N6M2F20highdate30.txt} ;
		\addlegendentry{Offline in (\ref{eq:simplfied_linear_program})};
		\addplot[line width=0.04cm, color = gray, mark = o, mark size = 4] table[x=lambda,y=AvgGanMN]{K6N6M2F20highdate30.txt} ;
		\addlegendentry{Scheme \cite{niesen2015coded}};
		
		
		\end{axis}
		\end{tikzpicture}
		\\ (a) \\
		\begin{tikzpicture}
		\begin{axis}[
		legend style={at={(0.73, 0.5)},anchor=north,nodes={scale=0.7, transform shape}},
		legend cell align = {left},
		xlabel = {\small $1/(F\lambda)$, Poisson Process parameter},
		ylabel = {\small Feasibility Probability},
		xlabel style={below},
		ylabel style={above},
		ylabel near ticks,
		yscale= 0.8,
		xmin = 0.0,
		xmax = 2.5,
		ymin = 0.0,
		ymax = 1.1,
		grid,
		]
		\addplot[line width=0.05cm, color = black, mark = square, mark size = 4] table[x=lambda,y=ProTtlOn]{K6N6M2F20lowdate26.txt} ;
		\addlegendentry{{\small Case I, low threshold}};
		\addplot[line width=0.05cm, color = brown, mark = triangle, mark size = 4] table[x=lambda,y=ProTtlOn]{K6N6M2F20highdate30.txt} ;
		\addlegendentry{{\small Case II, high threshold}};
		\addplot[line width=0.05cm, color = gray, mark = o, mark size = 4] table[x=lambda,y=ProTtlMN]{K6N6M2F20highdate30.txt} ;
		\addlegendentry{{\small All users \cite{niesen2015coded}}};
		\addplot[line width=0.05cm, color = gray, mark = diamond, mark size = 4] table[x=lambda,y=ProParMN]{K6N6M2F20highdate30.txt} ;
		\addlegendentry{{\small Recovery of a subfile in \cite{niesen2015coded}}};
		\end{axis}
		\end{tikzpicture}
		\\ (b)
	\end{tabular}
	

	\caption{\label{Fig:single_packetK6M2N6}{\small Decentralized placement scheme with deadlines for individual subfiles for $K=N=6$, $M=2$, and $F=20$: (a) average coding gain over all feasible offline problem instances, (b) feasibility probability of the online algorithm conditioned on feasibility of the offline problem. For the scheme in \cite{niesen2015coded} two probabilities are reported. The first one is the probability that all requests are satisfied, and the second one is the probability that a fixed request is satisfied (lines with circle and diamond marks respectively).
			At each trial, the cache content of each user is populated randomly and uniformly. In this simulation, we set $\eta_{0} = 0.4 - \frac{0.5}{\lambda }$  and $\eta_{0} = 0.8 - \frac{0.2}{\lambda }$ in Case I and Case II respectively.
		}
	}

\vspace{-0.25in}
\end{figure}
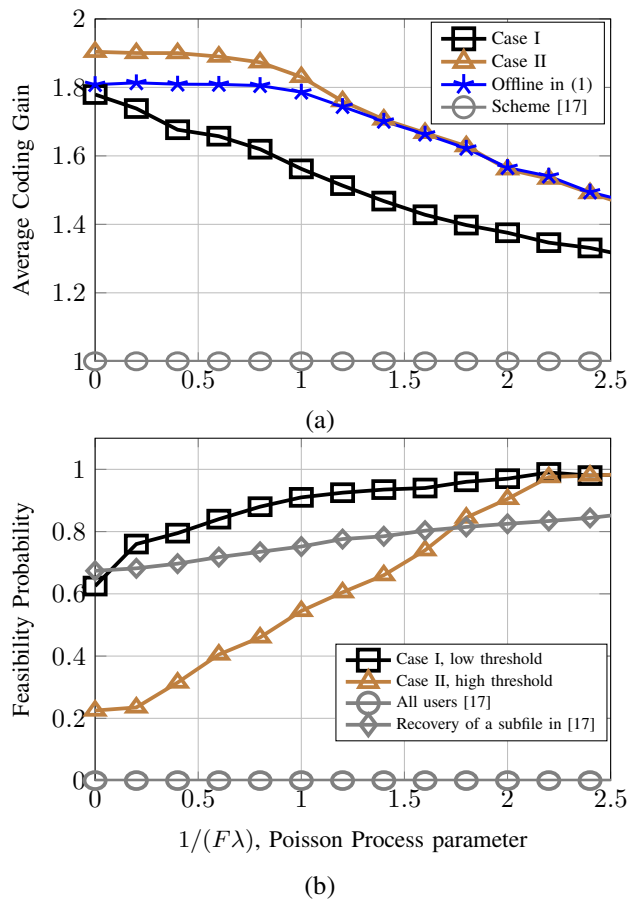
\vspace{-0.1in}
\subsection{Scenario where individual subfiles have deadlines}
The work of \cite{niesen2015coded} considers a situation where each subfile has its own deadline. This is inspired by applications such as video delivery over the Internet. We emphasize that this setting can be captured by our techniques. In particular, suppose that each user requests a set of subfiles from the server where the subfile requests arrive at different times and each subfile has a different deadline. In this case, we can treat each subfile request of user $i \in [K]$ as corresponding to a distinct virtual user whose cache content is the same as user $i$. However, the requests of the users are different. In this situation, each virtual user has precisely one missing subfile. Thus, the issue of coding over the corresponding subfiles does not arise.

Our setting is again one where $K=N=6$, $M=2$. Each file is subdivided into $F=20$ subfiles. Arrival times and deadlines are generated similar to the previous simulations with Poisson parameters $1 / (\lambda F)$ and the deadlines are randomly chosen uniformly from $[\Delta_{\min}, \Delta_{\max}]$ with $\Delta_{\min} = KMF/N$ and $\Delta_{\max} = KF$. Similar to the previous experiments we run $200$ trials and at each trial, the cache content of each user is populated randomly and uniformly among all placement schemes with cache of size $MF$ subfiles. Thus, different users might request different number of subfiles from the server. The only difference is that here each requested subfile has its own arrival time and deadline. The results are illustrated in Fig. \ref{Fig:single_packetK6M2N6}. It can be observed that our proposed approach provides significantly superior coding gain and feasibility probability as compared to the work of \cite{niesen2015coded}.



\vspace{-0.1in}
\section{Conclusions and Future Work}
\label{sec:conclusion}
In this work, we considered the asynchronous coded caching problem where user requests (with deadlines) arrive at the main server at different times. We considered both offline and online versions of this problem.
We demonstrated that under the assumption of all-but-one equations, the offline scenario can be optimally solved by a linear program (LP). Moreover, we presented a low-complexity solution to this LP based on dual decomposition. In contrast to the synchronous case and the offline scenario, we show that the online scenario requires coding across missing subfiles of a given user. Furthermore, we present an online algorithm that leverages offline LP in a recursive fashion. Extensive simulation results indicate that our proposed algorithm significantly outperforms prior algorithms.

Our online algorithm considers the situation where there is no knowledge about future request arrival times and file identities; this corresponds to a worst-case scenario. It would be interesting to consider cases where there is statistical information available on the arrival times and file popularity and/or algorithms where these can be learned and to investigate how this knowledge can be used to further improve the performance of the algorithm. For instance, it may be possible to design better placement schemes under this knowledge. Throughout the paper, we implicitly considered the case when different users request different files. It may be interesting to adapt our techniques for the case of repeated requests. 
\bibliographystyle{IEEETran}
\bibliography{coded_caching}

\begin{thebibliography}{10}
\providecommand{\url}[1]{#1}
\csname url@samestyle\endcsname
\providecommand{\newblock}{\relax}
\providecommand{\bibinfo}[2]{#2}
\providecommand{\BIBentrySTDinterwordspacing}{\spaceskip=0pt\relax}
\providecommand{\BIBentryALTinterwordstretchfactor}{4}
\providecommand{\BIBentryALTinterwordspacing}{\spaceskip=\fontdimen2\font plus
\BIBentryALTinterwordstretchfactor\fontdimen3\font minus
  \fontdimen4\font\relax}
\providecommand{\BIBforeignlanguage}[2]{{%
\expandafter\ifx\csname l@#1\endcsname\relax
\typeout{** WARNING: IEEEtran.bst: No hyphenation pattern has been}%
\typeout{** loaded for the language `#1'. Using the pattern for}%
\typeout{** the default language instead.}%
\else
\language=\csname l@#1\endcsname
\fi
#2}}
\providecommand{\BIBdecl}{\relax}
\BIBdecl

\bibitem{maddahN14}
M.~Maddah-Ali and U.~Niesen, ``Fundamental limits of caching,'' \emph{IEEE
  Trans. on Info. Th.}, vol.~60, no.~5, pp. 2856--2867, May 2014.

\bibitem{arbabjolfaei2018fundamentals}
F.~Arbabjolfaei, Y.-H. Kim \emph{et~al.}, ``Fundamentals of index coding,''
  \emph{Foundations and Trends{\textregistered} in Communications and
  Information Theory}, vol.~14, no. 3-4, pp. 163--346, 2018.

\bibitem{ghasemi2017asynch}
H.~Ghasemi and A.~Ramamoorthy, ``Asynchronous coded caching,'' \emph{IEEE Intl.
  Symp. on Info. Th.}, pp. 2438--2442, 2017.

\bibitem{rk}
R.~K. Ahuja, T.~L. Maganti, and J.~B. Orlin, \emph{Network Flows:Theory,
  Algorithms and Applications}.\hskip 1em plus 0.5em minus 0.4em\relax
  Prentice-Hall, 1993.

\bibitem{ghasemiR17_jnl}
H.~Ghasemi and A.~Ramamoorthy, ``Improved lower bounds for coded caching,''
  \emph{IEEE Trans. on Info. Th.}, vol.~63, no.~7, pp. 4388--4413, 2017.

\bibitem{yu2017exact}
Q.~Yu, M.~A. Maddah-Ali, and A.~S. Avestimehr, ``The exact rate-memory tradeoff
  for caching with uncoded prefetching,'' \emph{IEEE Trans. on Info. Th.},
  vol.~64, no.~2, pp. 1281--1296, 2017.

\bibitem{QYuMA19}
Q.~{Yu}, M.~A. {Maddah-Ali}, and A.~S. {Avestimehr}, ``Characterizing the
  rate-memory tradeoff in cache networks within a factor of 2,'' \emph{IEEE
  Trans. on Info. Th.}, vol.~65, no.~1, pp. 647--663, 2019.

\bibitem{TangR16}
L.~Tang and A.~Ramamoorthy, ``Coded caching for networks with the resolvability
  property,'' in \emph{IEEE Intl. Symp. on Info. Th.}, 2016.

\bibitem{wanJPT18}
K.~{Wan}, M.~{Ji}, P.~{Piantanida}, and D.~{Tuninetti}, ``Caching in
  combination networks: Novel multicast message generation and delivery by
  leveraging the network topology,'' in \emph{IEEE Intl. Conf. Comm.}, 2018,
  pp. 1--6.

\bibitem{naderializadeh2017fundamental}
N.~Naderializadeh, M.~A. Maddah-Ali, and A.~S. Avestimehr, ``Fundamental limits
  of cache-aided interference management,'' \emph{IEEE Trans. on Info. Th.},
  vol.~63, no.~5, pp. 3092--3107, 2017.

\bibitem{yanCTC17}
Q.~{Yan}, M.~{Cheng}, X.~{Tang}, and Q.~{Chen}, ``On the placement delivery
  array design for centralized coded caching scheme,'' \emph{IEEE Trans. on
  Info. Th.}, vol.~63, no.~9, pp. 5821--5833, 2017.

\bibitem{tang2018coded}
L.~Tang and A.~Ramamoorthy, ``Coded caching schemes with reduced
  subpacketization from linear block codes,'' \emph{IEEE Trans. on Info. Th.},
  vol.~64, no.~4, pp. 3099--3120, 2018.

\bibitem{lampirisE18}
E.~{Lampiris} and P.~{Elia}, ``Adding transmitters dramatically boosts
  coded-caching gains for finite file sizes,'' \emph{IEEE J. Select. Areas
  Comm.}, vol.~36, no.~6, pp. 1176--1188, 2018.

\bibitem{li2016fundamental}
S.~Li, M.~A. Maddah-Ali, Q.~Yu, and A.~S. Avestimehr, ``A fundamental tradeoff
  between computation and communication in distributed computing,'' \emph{IEEE
  Trans. on Info. Th.}, vol.~64, no.~1, pp. 109--128, 2017.

\bibitem{kiamari2017heterogeneous}
M.~Kiamari, C.~Wang, and A.~S. Avestimehr, ``On heterogeneous coded distributed
  computing,'' in \emph{IEEE Global Telecommunications Conference (GLOBECOM)},
  2017.

\bibitem{kostasR18}
K.~Konstantinidis and A.~Ramamoorthy, ``Leveraging coding techniques for
  speeding up distributed computing,'' in \emph{IEEE Global Telecommunications
  Conference (GLOBECOM)}, 2018.

\bibitem{niesen2015coded}
U.~Niesen and M.~A. Maddah-Ali, ``Coded caching for delay-sensitive content,''
  in \emph{IEEE Intl. Conf. Comm.}, 2015, pp. 5559--5564.

\bibitem{maddah2015decentralized}
M.~A. Maddah-Ali and U.~Niesen, ``Decentralized coded caching attains
  order-optimal memory-rate tradeoff,'' \emph{IEEE/ACM Trans. Netw.}, vol.~23,
  no.~4, pp. 1029--1040, 2015.

\bibitem{luCP18}
Y.~{Lu}, W.~{Chen}, and H.~V. {Poor}, ``Coded joint pushing and caching with
  asynchronous user requests,'' \emph{IEEE J. Select. Areas Comm.}, vol.~36,
  no.~8, pp. 1843--1856, 2018.

\bibitem{jiangHBZ19}
Y.~{Jiang}, W.~{Huang}, M.~{Bennis}, and F.~{Zheng}, ``Decentralized
  asynchronous coded caching design and performance analysis in fog radio
  access networks,'' \emph{IEEE Trans. on Mob. Comput.}, 2019 (to appear).

\bibitem{yang2018audience}
\BIBentryALTinterwordspacing
Q.~Yang, M.~M. Amiri, and D.~G{\"u}nd{\"u}z. Audience-retention-rate-aware
  caching and coded video delivery with asynchronous demands. [Online].
  Available: \url{https://arxiv.org/abs/1808.04835}
\BIBentrySTDinterwordspacing

\bibitem{vaidyaLP}
P.~M. Vaidya, ``{An algorithm for linear programming which requires
  $O(((m+n)n^2+(m+n)1.5 n)L)$ arithmetic operations},'' \emph{Math. Prog.},
  vol.~47, pp. 175--201, 1990.

\bibitem{lun2006minimum}
D.~S. Lun, N.~Ratnakar, M.~M{\'e}dard, R.~Koetter, D.~R. Karger, T.~Ho,
  E.~Ahmed, and F.~Zhao, ``Minimum-cost multicast over coded packet networks,''
  \emph{IEEE Trans. on Info. Th.}, vol.~52, no.~6, pp. 2608--2623, 2006.

\bibitem{boyd2004convex}
S.~Boyd and L.~Vandenberghe, \emph{Convex optimization}.\hskip 1em plus 0.5em
  minus 0.4em\relax Cambridge university press, 2004.

\bibitem{lemon}
\BIBentryALTinterwordspacing
LEMON. Library for efficient modeling and optimization in networks. [Online].
  Available: \url{http://lemon.cs.elte.hu}
\BIBentrySTDinterwordspacing

\bibitem{sherali1996recovery}
H.~D. Sherali and G.~Choi, ``{Recovery of primal solutions when using
  subgradient optimization methods to solve Lagrangian duals of linear
  programs},'' \emph{Operations Research Letters}, vol.~19, no.~3, pp.
  105--113, 1996.

\bibitem{ghasemiThesis}
\BIBentryALTinterwordspacing
H.~Ghasemi, ``Coded caching: Information theoretic bounds and asynchronism.''
  Ph.D. dissertation, Iowa State University, 2019. [Online]. Available:
  \url{http://www.ece.iastate.edu/~ghasemi/publication}
\BIBentrySTDinterwordspacing

\bibitem{kleinberg2006algorithm}
J.~Kleinberg and E.~Tardos, \emph{Algorithm design}, 2006.

\bibitem{asynch_sotware}
\BIBentryALTinterwordspacing
 [Online]. Available: \url{https://hooshanggh.github.io/Coded-Caching/}
\BIBentrySTDinterwordspacing

\bibitem{ho2006random}
T.~Ho, M.~Medard, R.~Koetter, D.~R. Karger, M.~Effros, J.~Shi, and B.~Leong,
  ``A random linear network coding approach to multicast,'' \emph{IEEE Trans.
  on Info. Th.}, vol.~52, no.~10, pp. 4413--4430, Oct 2006.

\end{thebibliography}

\vspace{-0.1in}
\appendix
\label{sec:appendix}

\subsection{Quadratic Projection and Primal Recovery in Dual Decomposition}
\label{app::primal_recovery}
For the projection of $\tilde{\gamma}^{(i)}_U(\ell,n)$ and $\tilde{\zeta}_\ell(n)$ to the constraint space
we simply set $\zeta_\ell(n) = \max\left(\tilde{\zeta}_\ell(n), \ 0 \right)$ and $\{\gamma_U^{(i)}(\ell,n),\ \forall \ i \in U \}$ is obtained via the following quadratic optimization. 
\begin{align}
\label{eq:Projection}
\min_{\{v^{(i)}_U: \ \forall \ i \in U\}} \sum_{i \in U} \left(v^{(i)}_U -  \tilde{\gamma}_U^{(i)}(\ell, n)\right)^2
~~\text{s.t.}~~  \sum_{i \in U} v^{(i)}_U = 1 .
\end{align}
In \cite[Appendix I]{lun2006minimum} an algorithm has been proposed to solve (\ref{eq:Projection}). This solution can be explained as follows.
For fixed $\ell \in [\beta]$ and for each $U \in \calU_\ell$, we sort $\tilde{\gamma}_U^{(i)}(\ell, n)$ so that $\tilde{\gamma}_U^{(i_1)}(\ell, n) \geq \ldots \geq \tilde{\gamma}_U^{(i_{|U|})}(\ell, n)$. We take $\hat{k}$ to be the minimum $k$ such that
\begin{align*}
\frac{1}{k} \left( 1-\sum_{j=1}^{k} \tilde{\gamma}_U^{(i_j)}(\ell, n) \right) \leq - \tilde{\gamma}_U^{(i_{k+1})}(\ell, n)
\end{align*}
or let $\hat{k} = |U|$ if such a $k$ doesn't exist. Then
$\gamma_U^{(i_j)}(\ell, n) = \tilde{\gamma}_U^{(i_j)}(\ell, n) + \frac{1-\sum_{l=1}^{\hat{k}} \tilde{\gamma}_U^{(i_l)}(\ell, n)}{\hat{k}}$ if $j\in [\hat{k}]$ and zero otherwise.

The initial setting for the dual variables is chosen as
$\gamma^{(i)}_U(\ell,0) = 1 / |U|$, for $i \in U, \ U \in \calU_\ell, \ \ell \in [\beta]$, and $\zeta_\ell = 0$ for $\ell \in [\beta]$. 

\noindent {\it Primal Recovery:} After solving the dual problem, the primal variables, i.e., $x_U(\ell, n)$'s, are recovered by the method of \cite{sherali1996recovery} whereby
\begin{align}
\label{eq:primal_recovery}
x_{U}(\ell, n) = \sum_{l=1}^n \mu_l(n) \left(\max_{i\in U} \ x^{(i)}_{U}(\ell, l)\right)
\end{align}
where $\mu_l(n)$'s are sequence of convex combination weights for each non-negative integer $n$, i.e. $\sum_{l=1}^n \mu_l(n) =1$ and $\mu_l(n) \geq 0$  for all $l=1,\ldots, n$. In \cite{lun2006minimum}, it has been shown that if the step size $\theta_n$ and convex combination weights $\mu_l(n)$ are chosen so that
\begin{itemize}
	\item $\eta_{l,n} \geq \eta_{l-1,n}$ for all $l=2,\ldots,n$ and $n=0,1,\ldots$,
	\item $\Delta_{\eta_n}^{\max} \rightarrow 0 $ as $n \leftarrow \infty$, and
	\item $\eta_{1,n} \rightarrow 0$ as $n \leftarrow \infty$ and $\eta_{n,n} \leq \delta$ for all $n=0,1,\ldots$ for some $\delta > 0$,
\end{itemize}
then $\{x_{U}(\ell, n)\}_{\ell \in [\beta],\ U\in \calU_l}$ is an optimal primal solution. Here $\eta_{l,n} = \frac{\mu_l(n)}{\theta_n}$ and
$\Delta_{\eta_n}^{\max} = \max_{l=2,\ldots, n} \{ \eta_{l,n} - \eta_{l-1,n} \}$.
Some sequences for $\theta_n$ and $\mu_l(n)$ that satisfy the above conditions has been proposed by \cite{lun2006minimum} . Among them we choose $\mu_l(n) = \frac{1}{n}$ and $\theta_n = n^{- \alpha}$
where $0 < \alpha < 1$. Then, the primal solution will be updated as,
\begin{align}
\label{eq:update_primal}
x_{U}(\ell, n+1) = \frac{n}{n+1}x_{U}(\ell, n) + \frac{\max_{i\in U} \ x^{(i)}_{U}(\ell, n)}{n+1}.
\end{align}
\subsection{Proof of Claim \ref{claim:online_feasibility}}
\label{proof_claim:online_feasibility}
\begin{IEEEproof}
	For simplicity, we prove the claim for $r=1$ and the proof for the general case follows directly.
	We will construct $x_U(\ell)$ and $y_{\{i,f\}}(U)$ variables for the offline LP from the decisions made in Algorithm \ref{Alg:Onlin_LP_rec}.
	Note that we update the set $\calX_{\text{off}}$ with the user groups chosen in Algorithm \ref{Alg:Onlin_LP_rec}.
	It is not difficult to verify that for any $\tilde{x}_U(\ell) \in \calX_{\text{off}}$ user group $U$ is a member of $\calU_{\ell}$. Moreover, the algorithm assigns integer values to $\tilde{x}_U(\ell)$. Now, for any $U \in \calU_{\ell}$ in (\ref{eq:simplfied_linear_program}), we set $x_U(\ell) = \tilde{x}_U(\ell)$ if $\tilde{x}_U(\ell) \in \calX_{\text{off}}$ and $x_U(\ell) = 0$ otherwise. Therefore, $x_U(\ell)$'s take integer values. Since at each time only one equation is transmitted in Algorithm \ref{Alg:Onlin_LP_rec}, the first condition $\sum_{U\in \calU_{\ell}} x_U(\ell) \leq |\Pi_{\ell}|$ holds for all $\ell \in [\beta]$. 

	For each $i\in [K]$ we define $[\tau_i, \tau_i+1)$ to be the last time slot that user $i$ benefits from the equation transmitted by the server. Clearly we have that $v_i(\tau_i+1) \geq |\Omega^{(i)}|$ otherwise Algorithm \ref{Alg:Onlin_LP_rec} will be infeasible at $\tau = T_i+\Delta_i$. We let $U_{i, \text{last}}$ to be the user group associated with this equation where $i \in U_{i, \text{last}}$.
	
	Note that Algorithm \ref{Alg:Onlin_LP_rec} tracks a set $\calU_{\text{sent}}(\tau)$ that contains all the user groups that have been used by the algorithm before time $\tau$.
	We let $\tilde{y}_{\{i,f\}}(U)$, $f \in \calF_{\{i,U\}}$ and $U \in \calU_{\text{sent}}(\tau_i)$ with $U \ni i$ be the solution of (\ref{eq:compute_eta}) when solving it for $w_{\{i,U_{i,\text{last}}\}}(\tau_i) $. Then, for each $U \in \cup_{\ell=1}^{\beta} \calU_{\ell}$ with $U \ni i$ and for each $f \in \calF_{\{i,U\}}$ we assign $y_{\{i,f\}}(U) = \tilde{y}_{\{i,f\}}(U)$ if $U \in \calU_{\text{sent}}(\tau_i)$ and $y_{\{i,f\}}(U)=0$ otherwise. We apply this assignment for all $i \in [K]$. Algorithm \ref{Alg:Onlin_LP_rec} assigns integer values to $z_U(\tau)$'s. From Remark \ref{rem::integrality_of_ys} it follows that there exists an integral solution for $\tilde{y}_{\{i,f\}}(U)$'s and consequently the $y_{\{i,f\}}(U)$'s as well. With these assignments, we now demonstrate that the second and third conditions in (\ref{eq:simplfied_linear_program}) hold.
	
	For the second condition we note that if $U \notin \calU_{\text{sent}}(\tau_i)$ then $y_{\{i,f\}}(U)=0$ and we have nothing to show. For $U \in \calU_{\text{sent}}(\tau_i)$ we have that $y_{\{i,f\}}(U) = \tilde{y}_{\{i,f\}}(U)$. Recall that $\tilde{y}_{\{i,f\}}(U)$ is the solution of (\ref{eq:compute_eta}) at time $\tau = \tau_i$.
	By the way that $z_U(\tau)$ has been updated in Algorithm \ref{Alg:Onlin_LP_rec}, we have $z_U(\tau) \leq z_U(T_{\max})$.
	Therefore, we have $z_U(\tau_i+1) \leq z_U(T_{\max}) = \sum_{\ell \in \calI_U} \tilde{x}_U(\ell)$ and from (\ref{eq:compute_eta}) for $w_{\{i,U_{i,\text{last}}\}}(\tau_i)$,
	\begin{align*}
	\sum_{f\in \calF_{\{i,U\}}} y_{\{i,f\}}(U) &= \sum_{f\in \calF_{\{i,U\}}} \tilde{y}_{\{i,f\}}(U)
	\leq \tilde{z}_U(\tau_i) = z_U(\tau_i+1) \\
	&\leq \sum_{\ell \in \calI_U} \tilde{x}_U(\ell)= \sum_{\ell \in \calI_U} {x}_U(\ell).	
	\end{align*}
	For the third condition, consider any user $i\in [K]$ and any $f \in \Omega^{(i)}$. Recalling the definition of $\tau_i$ and $w_{\{i,U_{i,\text{last}}\}}(\tau_i)$, we know that $w_{\{i,U_{i,\text{last}}\}}(\tau_i) = v_i(\tau_i+1) \geq |\Omega^{i}|$ which implies that in (\ref{eq:compute_eta}), we have
	\begin{align*}
	|\Omega^{(i)}| &\leq \sum_{U \in \calU_{\text{sent}(\tau_i)}, U \ni i} \sum_{f \in \calF_{\{i,U\}}} \tilde{y}_{\{i,f\}}(U)\\
	&=  \sum_{f\in \Omega^{(i)}} \sum_{U \in \calU_{\{ i, f\}} \cap \calU_{\text{sent}}(\tau_i)} \tilde{y}_{\{i,f\}}(U)
	\leq \sum_{f\in \Omega^{(i)}} 1 = |\Omega^{(i)}|,
	\end{align*}
	where the last inequality comes from the second constraint in (\ref{eq:compute_eta}). The middle equality holds by counting arguments for missing subfiles $f\in\Omega^{(i)}$ and user groups in $U \in \calU_{\text{sent}}(\tau_i)$.
	To verify this, consider a bipartite graph in which the left and right nodes correspond to $f \in \Omega^{(i)}$ and $U \in \calU_{\text{sent}}(\tau_i)$ with $U \ni i$ respectively. There is an edge between nodes corresponding to $f$ and $U$ if and only if $f \in \calF_{\{i,{U}\}}$. We let $\tilde{y}_{\{i,f\}}(U)$ to be the label of this edge. By the definition of $\calU_{\{ i, f\}}$ we know that $f \in \calF_{\{i,{U}\}}$ implies $U \in \calU_{\{i,f \}}$. Therefore, outgoing edges from the node corresponding to $f$ are the edges between $f$ and the nodes $U \in \calU_{\{i,f \}} \cap \calU_{\text{sent}}(\tau_i)$. Similarly, the outgoing edges between node $U \in \calU_{\text{sent}}(\tau_i)$ with $U\ni i$ are the edges between $U$ and $f \in \calF_{\{i,{U}\}}$. By counting $\tilde{y}_{\{i,f\}}(U)$ in two ways, from the left and right nodes, we have the required equality.
	Therefore, we have that $\sum_{U \in \calU_{\{ i, f\}} \cap \calU_{\text{sent}}(\tau_i)} \tilde{y}_{\{i,f\}}(U) = 1$ for any $f \in \Omega^{(i)}$. This further implies that $\sum_{U \in \calU_{\{ i, f\}} } {y}_{\{i,f\}}(U) = 1$ for all $f \in \Omega^{(i)}$ and completes the proof.	
\end{IEEEproof}

\subsection{Proof of Lemma \ref{lemma:zippleSchwart}}
\label{proof_lemma:zippleSchwart}
\begin{IEEEproof}
	For simplicity, in the discussion below we assume that $r = 1$. The proof for $r > 1$ follows in straightforward manner.	
	By the way that $\calM_i$ and $v_i(\tau)$ are updated in Algorithm \ref{Alg:Onlin_LP_rec}, we have $|\calM_i| = v_i(\tau)$ at each time $\tau$. Furthermore, $v_i(T_{\max}) = |\Omega^{(i)}|$ for all $i \in [K]$. Therefore, each user $i \in [K]$ benefits from $|\Omega^{(i)}|$ equations. For a $m \in \calM_i$, let $\bigoplus_{i \in U} \bigoplus_{f \in \calF_{\{i,U\}}}\alpha_{\{i, f, m\}} W_{\{d_i, f\}}$ represent the $m$-th equation (the dependence on index $j$ is suppressed since we assume that $r=1$). User $i \in U$ can recover $\bigoplus_{f \in \calF_{\{i,U\}}}\alpha_{\{i, f, m\}} W_{\{d_i, f\}}$ from this equation since the missing subfiles $W_{\{d_j,f'\}}$, for $f' \in\calF_{\{j,U\}}$ and $j \in U \setminus \{i\}$, exist in the cache of user $i$.
	
	For each user $i \in [K]$ we define matrix $\B_i \in \mathbb{F}^{|\Omega^{(i)}| \times |\Omega^{(i)}|}$ whose rows and columns correspond to equation numbers in $\calM_i$ and missing subfiles in $\Omega^{(i)}$ respectively. For $m \in \calM_i$, assume that $m$-th equation is associated with user group $U$, where $i \in U$. Then, the entry of $\B_i$ for the row and column corresponding to $m \in \calM_i$ and $f \in \Omega^{(i)}$ is $\alpha_{\{i, f, m\}}$ if $f \in \calF_{\{i,U\}}$ and zero otherwise. 
	Therefore, if matrix $\B_i$ is invertible then user $i$ can recover all the missing subfiles $W_{\{d_i,f\}}$, for $f\in \Omega^{(i)}$, from equations $\sum_{f \in \calF_{\{i,U\}}}\alpha_{\{i, f, m\}} W_{\{d_i, f\}}$ for $m\in \calM_i$. Thus, we need to show that the determinant of $\B_i$ is nonzero for all $i\in [K]$ with high probability.
	
	Towards this end, let $h_i( \{\alpha_{\{i, f, m\}}, \ f\in \Omega^{(i)} ,\ m\in \calM_i \})$ denote the determinant of $\B_i$; we treat the $\{\alpha_{\{i, f, m\}}, \ f\in \Omega^{(i)} ,\ m\in \calM_i \}$ as indeterminates at this point. Note that since Algorithm \ref{Alg:Onlin_LP_rec} did not return ``{\ttfamily INFEASIBLE}", we have a feasible integral solution for the corresponding offline LP ({\it cf.} Claim \ref{claim:online_feasibility}). Thus, there exists an interpretation of this solution ({\it cf.} Section \ref{subsec:LP_interper}) such that in each time slot, only one equation is transmitted, i.e., unlike a fractional solution, we do not need to potentially transmit multiple equations in the same time slot. This in turn implies that there is a setting for coefficients $\alpha_{\{i, f, m\}}$ with $\alpha_{\{i, f, m\}} \in \{0,1\}$ such that the multivariate polynomial $h_i$ evaluates to a non-zero value over $\mathbb{F}$, i.e., $h_i$ is not identically zero. This further implies that $h = \prod_{i\in[K]} h_i$ is not identically zero. Now, since each $\alpha_{\{i, f, m\}}$ appears only once in $\B_i$ thus its degree in polynomial $h_i$ is one. Also, $h_i$ is a polynomial of degree $|\Omega^{(i)}|\leq F$ thus $h$ is a polynomial of degree at most $KF$. Therefore, we can use Lemma 4 in \cite{ho2006random} to show that by choosing $\alpha_{\{i, f, m\}}$'s independently and uniformly at random from $\mathbb{F}$, the determinants of $\B_i$'s, $i\in [K]$, are nonzero with probability at least $\left(1 - \frac{1}{|\mathbb{F}|}\right)^{KF}$.
	
	When $r > 1$ we will need to split a missing subfile $W_{\{d_i, f\}}$ into $r$ packets and code over these as well. Thus, the corresponding system of equations will be of size $\mathbb{F}^{r|\Omega^{(i)}| \times r|\Omega^{(i)}|}$ leading to the bound $\left(1 - \frac{1}{|\mathbb{F}|}\right)^{r KF}$.
\end{IEEEproof}

\end{document}